\documentclass[12pt,reqno]{article}
\usepackage{times}

\usepackage{amsmath}
\usepackage{amsthm}
\usepackage{amsfonts}
\usepackage{amssymb}
\usepackage{epic}
\usepackage{eepic}
\usepackage{epsfig}

\usepackage{times, macros}
\usepackage{graphics,color}
\newcommand{\crd}{}

\theoremstyle{plain}

\theoremstyle{remark}

\newcommand{\sst}{\scriptscriptstyle}

\newcommand{\ot}{\otimes}
\newcommand{\ra}{\to}
\newcommand{\fr}[2]{{\textstyle \frac{#1}{#2} }}

\newcommand{\fsl}{{\mathfrak s}{\mathfrak l}}

\newcommand{\al}{\alpha}
\newcommand{\be}{\beta}
\newcommand{\ga}{\gamma}

\newcommand{\de}{\delta}

\newcommand{\ep}{\epsilon}
\newcommand{\la}{\lambda}

\newcommand{\si}{\sigma}

\newcommand{\vf}{\varphi}

\newcommand{\pa}{\partial}
\newcommand{\CA}{{\mathcal A}}
\newcommand{\CB}{{\mathcal B}}

\newcommand{\CD}{{\mathcal D}}
\newcommand{\CE}{{\mathcal E}}
\newcommand{\CF}{{\mathcal F}}

\newcommand{\CJ}{{\mathcal J}}

\newcommand{\CL}{{\mathcal L}}
\newcommand{\CM}{{\mathcal M}}

\newcommand{\CO}{{\mathcal O}}  
\newcommand{\CP}{{\mathcal P}}  
\newcommand{\CR}{{\mathcal R}}
\newcommand{\CS}{{\mathcal S}}
\newcommand{\CT}{{\mathcal T}}
\newcommand{\CU}{{\mathcal U}}

\newcommand{\CW}{{\mathcal W}}
\newcommand{\CX}{{\mathcal X}}
\newcommand{\CY}{{\mathcal Y}}
\newcommand{\CZ}{{\mathcal Z}}

\newcommand{\SH}{{\mathsf H}}

\newcommand{\SJ}{{\mathsf J}}

\newcommand{\fg}{{\mathfrak g}}
\renewcommand{\sf}{{\mathsf f}}
\newcommand{\spp}{{\mathsf p}}
\newcommand{\sx}{{\mathsf x}}
\newcommand{\hfg}{\hat{{\mathfrak g}}}

\newcommand{\1}{{\mathfrak 1}}
\newcommand{\2}{{\mathfrak 2}}
\newcommand{\3}{{\mathfrak 3}}

\newcommand{\BR}{{\mathbb R}}

\newcommand{\BC}{{\mathbb C}}

\newcommand{\BP}{{\mathbb P}}

\newcommand{\BZ}{{\mathbb Z}}

\newcommand{\rf}[1]{(\ref{#1})}

\begin{document}\thispagestyle{empty}
\title{Quantisation conditions of the quantum Hitchin system and the real geometric Langlands correspondence}
\author{J\"org Teschner}
\address{Department of Mathematics, \\
University of Hamburg, \\
Bundesstrasse 55,\\
20146 Hamburg, Germany,\\[1ex]
and:\\[1ex]
DESY theory, \\
Notkestrasse 85,\\
20607 Hamburg,
Germany}
\maketitle
\section{Introduction}

\subsection{Motivations}

-- {\it Program of Nekrasov-Shatashvili \cite{NS}, relations to gauge theory}

An important  motivation comes from the program initiated by Nekrasov and
Shatashvili investigating relations between supersymmetric field theories and quantum integrable 
models. An interesting family of examples to which this program can be applied is provided by
a class of four-dimensional $\mathcal{N}=2$-supersymmetric field theories associated to 
the choice of a pair $(C,\fg)$ consisting of a (possibly punctured) Riemann surface  $C$
and a Lie-algebra $\fg$ of ADE-type \cite{Ga09,GMN}. The integrable models relevant for this class of theories 
are known \cite{GMN} to be the Hitchin systems \cite{Hi87b}. Regularising the 
supersymmetric field theories by means of the so-called Omega-deformation leads to the 
quantisation of the corresponding integrable models \cite{NS,NW,NRS}. 

-- {\it Quantum integrable systems}

Many quantum integrable models can be solved by the Bethe ansatz method.
Whenever the Bethe ansatz is applicable, it 
is often useful to formulate the Bethe ansatz equations representing the quantisation conditions 
in terms of a single, model-dependent function $\CY(\mathbf{a},\mathbf{t})$ called Yang's function
\cite{YY}. This function depends on two types of variables, $\mathbf{a}=(a_1,\dots,a_d)$ and 
$\mathbf{t}=(\tau_1,\dots,\tau_{d'})$. The parameters
$\mathbf{t}$ are parameters of the commuting Hamiltonians, in the context of spin chains often called
inhomogeneity parameters.  The variables $\mathbf{a}$ are auxiliary, allowing us to represent the 
Bethe Ansatz equations in the form
\begin{equation}\label{q-cond-a}
\frac{\partial}{\partial a_k}\CY(\mathbf{a},\mathbf{t})\,=\,2\pi i n_k,\qquad 
k=1,\dots,d.
\end{equation} 
In a non-degenerate situation equation \rf{q-cond-a} 
has a unique solution $\mathbf{a}=\mathbf{a}_{\rm cr}(\mathbf{n})$ for given  integers $\mathbf{n}=(n_1,\dots,n_d)$.
The eigenvalues $E_r$ of a subset of 
the commuting conserved quantities  $\mathsf{H}_r$, $r=1,\dots,d'$,
can the be obtained from $\CY(\mathbf{a},\mathbf{t})$ by taking the derivatives
\begin{equation}
E_r\,=\, \frac{\pa}{\pa \tau_r}\CY(\mathbf{a},\mathbf{t})\Big|_{\mathbf{a}=\mathbf{a}_{\rm cr}(\mathbf{n})}\,.
\end{equation}

Beyond the class of  quantum integrable systems soluble by Bethe ansatz techniques, there exists a 
large class of models where such techniques fail.
One important outcome of the Nekrasov-Shatashvili program is strong evidence for the proposal made
in \cite{NS}  
that the quantisation conditions in large classes of 
integrable models which can not be solved by the Bethe ansatz method can nevertheless be 
described in terms of suitable Yang's functions.

However, for the models studied in this paper it will turn out  that another 
type of condition formulated in terms of a single function $\CY(\mathbf{a},\mathbf{t})$ 
is appropriate. In general it is not a priori obvious
which type of  condition is appropriate for a given model.
The scheme of Nekrasov-Shatashvili will  be efficient for the solution of 
quantum integrable systems only if one knows exactly how a given quantisation 
condition is represented in terms of the Yang's function $\CY(\mathbf{a},\mathbf{t})$. Answering this question
for interesting integrable models may lead into fairly profound mathematical problems, 
as will be illustrated by the examples studied in this paper.

-- {\it Geometric Langlands program}

The geometric Langlands correspondence is often loosely formulated as a correspondence which assigns
$\mathcal{D}$-modules on ${\rm Bun}_G$ to ${}^LG$-local systems on a Riemann surface $C$, 
see \cite{Flect} for a review of the aspects relevant here.
${}^LG$ is the Langlands dual group of a simple complex Lie group $G$.
Most interesting for us is the special case considered in the original work of Beilinson and Drinfeld
where the ${}^LG$-local systems are {\it opers}, pairs $(\mathcal{E},\nabla')$
in which  $\nabla'$ is gauge-equivalent to a certain standard form.
The space of opers 
forms a Lagrangian subspace in the moduli space of all local systems.
The corresponding 
$\mathcal{D}$-modules on ${\rm Bun}_G$ can be described more concretely as systems of 
partial differential equations taking the form of eigenvalue equations 
$\SH_r f= E_rf$ for a family of differential operators $\SH_r$ on ${\rm Bun}_G$  quantising 
the Hamiltonians of Hitchin's integrable system. The oper corresponding to such a $\CD$-module 
in the geometric Langlands correspondence is the geometric object encoding the eigenvalues $E_r$.

-- {\it Relations to conformal field theory}

This paper is part of a larger program outlined in \cite{T10,BT} on the relations between the quantisation of the Hitchin
system, supersymmetric field theories, 
conformal field theory, and the geometric Langlands program. Some of these relations 
will be briefly described at the end of this paper. 




\subsection{Main results}

We are going to propose a natural quantisation condition for the Hitchin system, and explain how it
can be reformulated in terms of a function $\CY(\mathbf{a},\mathbf{t})$. The function 
$\CY(\mathbf{a},\mathbf{t})$ relevant for this task
is found to be the generating function for the variety of opers within the 
space of all local systems as predicted in \cite{NRS,T10}. However, the condition on $\CY$ expressing the quantisation 
condition turns out to be different from the types of conditions
considered in \cite{NS}. 
Our derivation is essentially complete for Hitchin systems associated to the Lie algebra $\fsl_2$
in genus $0$ and $1$, which may be called the Gaudin and elliptic Calogero-Moser models
assciated to the group $\mathrm{SL}(2,\BC)$. It reduces to a conjecture of E. Frenkel \cite{Ficmp}
for $g>1$, as will be discussed below. 

Reformulating the quantisation conditions in terms of $\CY$ 
can be done using the Separation of Variables (SOV) 
method pioneered by Sklyanin \cite{Skl}. This method may be seen as a more concrete 
procedure to construct the geometric Langlands correspondence relating opers to 
$\CD$-modules (eigenvalue equations), 
as was pointed out in \cite{Ficmp}. In our case it will be found that the SOV method relates 
single-valued solutions of the eigenvalue equations to opers having real holonomy.
This problem is closely related to the classification of  projective structures on $C$ with 
real holonomy which has been studied in \cite{Go87}. Using complex Fenchel-Nielsen
coordinates we will reformulate this description in terms of the generating 
function for the variety of opers.

From the point of view of the geometric Langlands correspondence we obtain a 
correspondence between opers with real holonomy
and $\CD$-modules admitting single-valued solutions. 
We expect that a generalisation to more general local systems with real holonomy will
exist. We propose to call such correspondences the real geometric Langlands correspondence.


\section{Separation of variables for the classical Hitchin integrable system}\label{class}

\subsection{Integrability and special geometry}\label{special}

A complex symplectic manifold $\CM$ with holomorphic symplectic form 
$\Omega$ is called an algebraic integrable system if it can be described as
a Lagrangian torus fibration $\pi:\CM\ra\CB$ with fibres being principally polarised abelian varieties. 
Algebraic integrability is equivalent to the fact that the base $\CB$ is a special K\"ahler manifold
satisfying certain integrality conditions \cite{Fd}.

These connections 
may be reformulated conveniently in terms of  a covering of $\CM$ with 
local charts carrying action-angle  
coordinates consisting of a tuple $\mathbf{a}=(a^1,\dots,a^d)$ of coordinates for the base
$\CB$, and  complex coordinates $\mathbf{z}=(z_1,\dots,z_d)$ for the torus fibres 
$\Theta_b=\BC^d/(\BZ^d+\tau_b\cdot\BZ^d)$, $b\in\CB$, such 
that
\begin{equation}
\Omega=\sum_{r=1}^d da^r\wedge dz_r.
\end{equation}
The transformation $\mathbf{z}^{\rm\sst D}:=\tau^{-1}_b\cdot \mathbf{z}$ gives an
equivalent representation of the torus fibres $\Theta_b$. It can be extended to a canonical transformation
$(\mathbf{a},\mathbf{z})\ra (\mathbf{a}^{\rm\sst D},\mathbf{z}^{\rm\sst D})$ by introducing coordinates 
$a_s^{\rm\sst D}$ satisfying $\frac{\pa}{\pa a^r}a_s^{\rm\sst D}=\tau_{rs}$. As $\tau_{rs}=\tau_{sr}$,
there exists a potential $\CF(\mathbf{a})$ allowing us to represent $a_r^{\rm\sst D}$ in the form
$a_r^{\rm\sst D}=\frac{\pa}{\pa a^r}\CF(\mathbf{a})$. It follows that
\begin{equation}
\Omega=\sum_{r=1}^d da^{\rm\sst D}_r\wedge dz_{\rm\sst D}^r.
\end{equation}

One may equivalently represent  $\Theta_b$ 
as {\it real} torus $\BR^{2d}/\BZ^{2d}$
using the coordinates $(\mathbf{w},\mathbf{w}_{\rm\sst D})$,
$\mathbf{w}=(w_1,\dots,w_d)$,  $\mathbf{w}_{\rm\sst D}=(w^1_{\rm\sst D},\dots,w^d_{\rm\sst D})$ such that 
$\mathbf{z}=\mathbf{w}+\tau\cdot \mathbf{w}_{\rm\sst D}$.
There exists a corresponding set of real action variables $(\mathbf{b},\mathbf{b}_{\rm\sst D})$ such that
\begin{equation}
\mathrm{Re}(\Omega)=\sum_{r=1}^d (db^r\wedge dw_r+db^{\rm\sst D}_r\wedge dw_{\rm\sst D}^r).
\end{equation}
The real action variables $(\mathbf{b},\mathbf{b}_{\rm\sst D})$ are simply the real parts of 
$(\mathbf{a},\mathbf{a}_{\rm\sst D})$.
The coordinates above are only locally defined, in general. Different 
sets of coordinates are related by $\mathrm{Sp}(2d,\BZ)$-transformations acting in the
standard fashion on the vectors $(\mathbf{w},\mathbf{w}_{\rm\sst D})$.

\subsection{Integrability of the Hitchin system}

The phase space $\CM_{\rm \sst H}(C)$ of the Hitchin system \cite{Hi87b} for $G=GL(2)$  on
a Riemann surface $C$ with genus $g>1$ 
is the moduli space of stable pairs $(\CE,\vf)$, where $\CE$ is a
holomorphic rank 2 vector bundle, and $\vf\in H^0(C,{\rm End}(\CE)\otimes K_C)$
is called the Higgs field, modulo  gauge transformations.  
There is a natural stability 
condition for the pairs $(\CE,\vf)$ allowing certain unstable bundles $\CE$.
The open dense subset of 
$\CM_{\rm \sst H}(C)$ consisting of pairs $(\CE,\vf)$ with stable bundles $\CE$ is isomorphic to the 
cotangent bundle $T^*\mathrm{Bun}_G(C)$. The moduli space $\CM_{\rm \sst H}(C)$ carries a natural 
holomorphic symplectic structure  restricting to the canonical symplectic structure 
on the dense open subset $T^*\mathrm{Bun}_G(C)$.
Considering bundles $\CE$ with 
fixed determinant and Higgs fields $\vf$ with vanishing trace 
allows one to describe the Hitchin system for $G=SL(2)$ in a similar way.

The complete integrability of the Hitchin system is demonstrated using
the so-called Hitchin map, in our case mapping a pair $(\CE,\vf)$ to 
the coefficients $(\vartheta_1,\vartheta_2)$ of the characteristic polynomial 
$\mathrm{det}(v\,\mathrm{id}-\vf(u))=v^2-\vartheta_1v+\vartheta_2$. 
The coefficients $(\vartheta_1,\vartheta_2)$ can be identified with elements
of  the vector space $\CB=H^0(C,K)\oplus H^0(C,K^2)$.
Fixing bases $\{\rho_1,\dots,\rho_g\}$ and $\{q_1,\dots,q_{3g-3}\}$  for $H^0(C,K)$ and $H^0(C,K^2)$, 
respectively, allows us to define the Hamiltonians of the 
Hitchin system to be the coefficients in the  expansions
${\rm tr}(\vf(u))=\sum_{i=1}^g \rho_i h_i$ and   ${\rm tr}(\vf^2(u))=\sum_{r=1}^{3g-3} q_r H_r$.
They form a maximal set of Poisson-commuting globally defined functions 
on $\CM_{\rm \sst H}(C)$. The Hitchin fibres $\Theta_{b}$ 
are the subvarieties of $\CM_{\rm \sst H}(C)$ associated to a point $b\in \CB$.

In order to see that generic  fibres $\Theta_{b}$ can be represented as abelian 
varieties (complex tori), one may first
define the spectral curve $\Sigma$ as
\begin{equation}\
\Sigma\,=\,\big\{\,(u,v)\in T^*C\,;\,\mathrm{det}(v\,\mathrm{id}-\vf)=0\,\big\}\,.
\end{equation}
To each pair  $(\CE,\vf)$ let us then associate a line bundle $L$ on $\Sigma$,
the bundle with fibres being the eigenlines of $\vf$ for a given eigenvalue
$v$, defining a map from $(\CE,\vf)$ to the pair $(\Sigma,L)$.
Conversely, given a pair $(\Sigma,L)$, where $\Sigma\subset T^*C$
is a double cover of $C$, and $L$ a holomorphic line bundle on
$\Sigma$, one can recover $(\CE,\vf)$ via
\begin{equation}
(\CE,\vf)\,:=\,\big(\,\pi_*(L)\,,\,
\pi_*(v)\,\big)\,,
\end{equation}
where $\pi$ is the covering map $\Sigma\ra C$, and $\pi_*$ is the direct
image. In this way we may  identify the Hitchin fibres $\Theta_{b}$  with the 
Jacobian of $\Sigma$ parameterising 
the choices of the line bundles $L$. 
This is how the space $\CM_{\rm \sst H}(C)$ gets described as torus fibration 
with the fibre over a point $b\in\CB$ being the Jacobian. 

For the case of $G=SL(2)$ one needs to impose the condition that the bundle $\CE$ has trivial determinant.
The Jacobian is then replaced by the so-called Prym variety parameterising line bundles $L$ such that
$\mathrm{det}(\pi_*(L))\simeq \CO$.

It can furthermore be shown that the dynamics of the Hitchin system generated by the Hamiltonians 
with respect to the natural symplectic structure  gets linear on the torus 
fibres \cite{Hi87b}, completing the proof of the complete integrability of the Hitchin system.

\subsection{Algebraic integrability of Jacobian fibrations}

Algebraic integrability is realised in a canonical fashion in terms of 
Jacobian or Prym fibrations of spectral curves. Indeed, given a spectral curve
$\Sigma$, let us pick 
a canonical 
basis for the first homology of $\Sigma$, represented by mutually 
nonintersecting sets of cycles $\al_1,\dots,\al_h$ and 
$\be_1,\dots,\be_h$ satisfying $\al_r\cdot\be_s=\de_{r,s}$, where $h=4g-3$ is the genus of $\Sigma$.
A basic role is played by the periods
\begin{equation}\label{periods}
a^r=\int_{\al_r}\la,\qquad a_r^{\rm\sst D}=\int_{\be_r}\la.
\end{equation}
of the canonical differential $\la=vdu$ on $\Sigma$. 
The derivatives $\omega_r=\pa_{a^r}\la$ give a basis for the 
space of abelian differentials 
normalised as
$
\de_{r,s}=\int_{\al_r}\omega_s, 
$
The torus fibres may then be represented as $\Theta_{\mathbf{E}}=\BC^h/(\BZ^h+\tau\cdot\BZ^h)$,
with period matrix $\tau$ having matrix elements
$
\tau_{rs}=\int_{\be_s}\omega_r.
$
The Riemann bilinear relations give $\tau_{rs}=\tau_{sr}$. It follows that 
there exists a function $\CF(\mathbf{a})$ giving the dual periods $a_r^{\rm\sst D}$ as 
$a_r^{\rm\sst D}=\pa_{a^r}\CF(\mathbf{a})$.



When the integrable structure is represented in terms of a torus fibration over families of 
spectral curves which are branched coverings of an underlying curve $C$,
one may alternatively represent the integrable structure in terms of a symmetric 
product $(T^*C)^{[h]}$ of the cotangent bundle of $C$. This relation is essentially canonical
and most easily described when the torus fibres are the Jacobians of $\Sigma$.
The Abel map from divisors $\hat{\mathbb{D}}_{\hat{\mathbf{u}}}=\sum_{r=1}^h \hat{u}_r$ on $\Sigma$ to the Jacobian,
\begin{equation}
z_s(\mathbf{a},\mathbf{u})=\sum_{r=1}^h \int^{\hat{u}_r}\omega_s. 
\end{equation}
can be inverted (Jacobi inversion problem), defining a divisor 
$\mathbb{D}_{\mathbf{u}}=\sum_{r=1}^h {u}_r$ on $C$ by projection.
The locally defined function  
\begin{equation}
\CX(\mathbf{a},\mathbf{u})=\sum_{r=1}^h \int^{\hat{u}_r}\la,
\end{equation}
is a generating function for the change of variables from $(\mathbf{a},\mathbf{z})$ to 
$(\mathbf{v},\mathbf{u})$, 
\begin{equation}\label{Xgenfct}
\frac{\pa}{\pa a_r}
\CX(\mathbf{a},\mathbf{u})=z_r,\qquad 
\frac{\pa}{\pa u_r}
\CX(\mathbf{a},\mathbf{u})=v_r.
\end{equation}
It follows from the existence of the generating function $\CX(\mathbf{a},\mathbf{u})$ that
the coordinates $(\mathbf{v},\mathbf{u})$ are Darboux coordinates.
Note that the points $(u_k,v_k)\in T^*C$ with $v_k=v_k(\mathbf{a},\mathbf{u})$ 
defined in \rf{Xgenfct} automatically satisfy 
\begin{equation}
v_k^2-\mathrm{tr}(\vf(u_k))+\mathrm{tr}(\vf^2(u_k))=0,\quad \Leftrightarrow\quad(u_k,v_k)\in\Sigma,
\end{equation}
for $k\in 1,\dots,{\crd h}$.
A detailed explanation of the modifications of the Abel map that are necessary in the cases where the torus 
fibres are Prym varieties can be found in \cite{Wi15}. Only the subspace of $H_1(\Sigma)$ which is odd under 
the exchange of sheets is relevant in this case, reducing the number of relevant variables from $h$ to 
$d=3g-3$.

The representation in terms of the symmetric 
product $(T^*C)^{[h]}$ will be called Separation of Variables (SOV) representation.
We conclude that a SOV representation  exists for the classical theory whenever there is a description in terms
of pairs $(\Sigma,L)$ as introduced above.

\subsection{Separation of variables}

It may be necessary to describe the passage from the original description in terms of 
pairs $(\CE,\vf)$ to either one of the two descriptions making the integrable structure manifest 
more explicitly.
This requires 
constructing sections $\chi$ of the line bundle $L$ as 
families of eigenvectors of the Higgs-field $\vf$. The divisor $D_{\mathbf{u}}$ will be identified 
with the divisor of zeros of $\chi$ \cite{Hurt,GNR}.

To begin with, we need to represent the pairs 
$(\CE,\vf)$ more concretely. This can be done
by representing the bundles $\CE$ as extensions,
\begin{equation}\label{EXT}
0\longrightarrow \CL'\longrightarrow \CE\longrightarrow \CL''\longrightarrow 0\,.
\end{equation}
Describing such extensions by means of a covering $\CU_\imath$ of $C$ and
transition functions $\CE_{\imath\jmath}$ between patches $\CU_{\imath}$ and
$\CU_{\jmath}$, one may assume that all $\CE_{\imath\jmath}$ are upper triangular,
\begin{equation}\label{transfct}
\CE_{\imath\jmath}\,=\,\bigg(\begin{matrix} \CL_{\imath\jmath}' & 0\\
0 & \CL_{\imath\jmath}''\end{matrix}\bigg)\bigg(\begin{matrix} 1 & \CE_{\imath\jmath}'\\
0 & 1\end{matrix}\bigg)\,.
\end{equation}
This implies that the lower left matrix element $\vf_-(y)$ of $\vf$  is a section of the line bundle $\CL\otimes K_C$,
with $K_C$ being the canonical line bundle and $\CL=(\CL')^{-1}\otimes\CL''$. 
Without loss of generality one may assume $\CL'=\CO$, $\CL''=\CL$, as can always by reached by 
tensoring $\CE$ with a line bundle.
Any holomorphic bundle can be represented as an extension \rf{EXT}. 
At least part of the moduli of the bundle $\CE$ can be represented in terms of 
extension classes in $\mathbb{P}H^1(\CL^{-1})$. Since ${\rm dim}H^1(\CL^{-1})=g-1+\mathrm{deg}(\CL)$ 
this suffices to represent all moduli of ${\rm Bun}_{\mathrm{SL}(2)}$ if $\mathrm{deg}(\CL)> 2g-2$. To simplify the 
discussion we shall assume $\mathrm{deg}(\CL)=2g-1$ in the following.

The matrix elements $\vf_-$ of $\vf$ represent elements of
the vector space  $H^0(C,\CL\ot K_C)$ dual to $H^1(\CL^{-1})$ by Serre duality.
The  eigenvectors of $\vf=\big( \begin{smallmatrix} \vf_0' & \vf_+ \\ \vf_- & \vf_0''
\end{smallmatrix}\big)$,
\begin{equation}
\chi=
\left(\begin{matrix} v-\vf_0'' \\ \vf_- \end{matrix}\right).
\end{equation}
vanish at the zeros of $v-\vf_0''$ which project to the $4g-3$ zeros $\mathbf{u}=(u_1,\dots,u_{h})$
of $\vf_-$ on $C$.  The degree  $4g-3$ line bundle $L=\CO(\hat{\mathbb{D}}_{\hat{\mathbf{u}}})$ associated to the
divisor $\hat{\mathbb{D}}_{\hat{\mathbf{u}}}=\sum_{r=1}^h\hat{u}_r$ represents the point in the Jacobian of $\Sigma$
associated to $(\CE,\vf)$.
We thereby obtain the relation between pairs $(\CE,\vf)$,  where $\CE$ is
represented as extension of the form \rf{EXT},  and the tuples of 
points $(\mathbf{u}, \mathbf{v})$ in 
$(T^*C)^{[h]}$ introduced above: $\mathbf{u}=(u_1,\dots,u_{h})$ is the 
collection of zeros of $\vf_-$, while $\mathbf{v}=(v_1,\dots,v_{h})$ is defined by 
setting $v_k=\vf_0''(u_k)$, $k=1,\dots,h$. 

In order to treat the  case of the $G=SL(2)$ Hitchin system one may consider the 
 line bundle $\CL\simeq \mathrm{det}(\CE)$ as fixed, which imposes $g$ constraints on the 
 positions of the $u_1,\dots,u_{h}$.
We furthermore have $\vf_0'=-\vf_0''\equiv\vf_0$. Let $\si$ be the sheet involution.
The degree zero line bundle $L=\CO(\hat{\mathbb{D}})$ associated to the
divisor $\hat{\mathbb{D}}=\sum_{r=1}^h(\hat{u}_r-\si(\hat{u}_r))$ representing the point in the 
Prym variety of $\Sigma$
associated to $(\CE,\vf)$
has lines generated by
 \begin{equation}
\chi=\frac{1}{v-\vf_0}
\left(\begin{matrix} v+\vf_0 \\ \vf_- \end{matrix}\right).
\end{equation}

Variants of this type of representation can be used to  parameterise the pairs
$(\CE,\vf)$, and to describe the change of variables defining the tuples $(\mathbf{u}, \mathbf{v})$, 
much more 
explicitly \cite{Kr01}. 


\subsection{Punctures}

It is possible to generalise the set-up by allowing $n$ marked points on $C$. In the presence
of marked points one may also consider surfaces of genus $0$ or $1$. The 
resulting versions of the Hitchin integrable systems turn out to be related to 
the integrable models known as Gaudin model ($g=0$), or the elliptic
Calogero-Moser model ($g=1$).
We will use the the example of the Gaudin model as guidance for the 
quantisation of the picture outlined above. The necessary ingredients 
will have clear analogs in this case, suggesting a path for the treatment of the general case.
To this aim let us explain how the separation
of variables is realised in this case. 

The description of $\CE$ as an extension amounts to a description in terms of a cover of $\BP^1$ of the 
form
$\{\BP^1\setminus\{z_1,\dots,z_n\},D_1,\dots,D_n\}$, where $D_1,\dots,D_n$ are small mutually non-intersecting
discs around $z_1,\dots,z_n$, with transition functions on $A_r=D_r\setminus \{z_r\}$ being 
of the form $\CE_r=\big(\begin{smallmatrix} 1 & x_r \\ 0 & 1\end{smallmatrix}\big)$. Assuming that 
$\vf$ has a regular singularity of the form $\frac{1}{y-z_r}
\big(\begin{smallmatrix} l_r & 0 \\ p_r & -l_r\end{smallmatrix}\big)$  at $z_r$ it follows that
\begin{equation}
\vf(y)=\sum_{r=1}^n\frac{\vf_r}{y-z_r}, \quad
\vf_r=
\CE_r^{}\cdot\bigg(\begin{matrix} l_r & 0 \\ p_r & -l_r\end{matrix}\bigg)\cdot\CE_r^{-1}
=\bigg(\begin{matrix} x_rp_r+l_r & x_r^2p_r+2l_rx_r \\ p_r & -l_r-x_rp_r\end{matrix}\bigg).
\end{equation}
Regularity of $\vf$ at infinity imposes three constraints
\begin{equation}\label{reginfty}
\sum_{r=1}^n x_r^{k+1}p_r+l_r(k+1)x_r^k=0,\qquad k=-1,0,1.
\end{equation}
Identifying $x_r$ with a coordinate on $\BP^1$, and $p_r$ with a coordinate on the cotangent 
fibre of $\BP^1$ allows us to describe $\CM_{\rm H}(C_{0,n})$ as symplectic reduction of 
$(T^*\BP^1)^n$ by the constraints \rf{reginfty}. To this aim one needs to identify 
points of $(T^*\BP^1)^n$ related by the Hamiltonian flows generated by the constraints.
These flows generate the group $G=\mathrm{SL}(2)$ acting on the variables $x_r$ as 
M\"obius transformations $x_r\ra \frac{ax_r+b}{c x_r+d}$.
The quotient $(T^*\BP^1)^n/\!\!/G$ may be 
represented by fixing a slice $x_n=\infty$, $x_{n-1}=1$ and $x_{n-2}=0$ and
using \rf{reginfty} to express $p_n$, $p_{n-1}$ and $p_{n-2}$ in terms of the 
remaining variables. This forces us to send $p_n\ra0$ such that 
$x_np_n+2l_n=0$.

The 
Hamiltonians of this integrable model are defined as the free parameters specifying 
the quadratic differential $\mathrm{tr}(\vf^2)$, which can now be represented
explicitly as 
\begin{equation}
\mathrm{tr}(\vf^2(y))=\sum_{r=1}^n\left(\frac{l_r^2}{(y-z_r)^2}+\frac{H_r}{y-z_r}\right).
\end{equation}

The change of variables $(\mathbf{x},\mathbf{p})\ra (\mathbf{u},\mathbf{v},u_0)$ defined by
\begin{equation}
\vf_-(y)=
\sum_{r=1}^{n-1}\frac{p_r}{y-z_r}=u_0\frac{\prod_{k=1}^{n-3}(y-u_k)}{\prod_{r=1}^{n-1}(y-z_r)},
\qquad v_r=\vf_0(u_r),
\end{equation}
gives the isomorphism $\CM_H(C_{0,n})\simeq (T^*C_{0,n})^{[n-3]}$ defined by the 
SOV method.

\section{Quantisation of Hitchin's integrable system}

We will now present an overview of known results on the quantisation of the Hitchin system.
Starting with the genus zero case  we will introduce a variant of the Gaudin model related
to the non-compact group $\mathrm{SL}(2,\BC)$. Known results on the quantisation 
of Hitchin's Hamiltonians in $g>1$ and their relation to the geometric Langlands correspondence 
are re-interpreted from the point of view of this paper in the following subsection.

\subsection{Genus zero -- the $\mathrm{SL}(2,\BC)$ Gaudin model}

The quantisation of the Gaudin model is fairly simple 
on a purely algebraic level. It starts by
turning the algebra of functions on $(T^*\mathbb{P}^1)^n$ with  generators $p_r$, $x_r$, 
into a non-commutative algebra with generators $\spp_r$, $\sx_r$, 
$r=1,\dots,n$,  satisfying the relations $[\spp_r,\sx_s]=\ep_1\de_{rs}$, $[\spp_r,\spp_s]=0$,
$[\sx_r,\sx_s]=0$. The matrix elements $\vf_r^a$, $a=-,0,+$ 
of the residues $\vf_r$ of $\vf$ get replaced by the generators of the Lie algebra
$\fsl_2$ for all $r=1,\dots,n$. The quantised algebra of functions $\mathfrak{A}_n$ on $(T^*\mathbb{P}^1)^n$
thereby gets identified with the direct sum of $n$ copies of the Lie algebra $\fsl_2$.

When we are discussing the quantisation of a phase space with complex coordinates 
it is also natural to consider the conjugate algebra $\bar{\CA}_n$ obtained by
quantisation of the complex conjugate coordinates $\bar{p}_r$, $\bar{x}_r$. The generators
of $\bar{\CA}_n$ will be denoted as $\bar{\spp}_r$, $\bar{\sx}_r$, $r=1,\dots,n$.

Recall that we had 
represented  $\CM_{\rm H}(C_{0,n})$ as symplectic quotient of $(T^*\mathbb{P}^1)^n$
by the three constraints \rf{reginfty}. The constraints become quantised to the ``diagonal''
$\fsl_2$ embedded into the direct sum of $n$ copies of $\fsl_2$ in the usual way. 
It is natural to define the quantised algebra $\CA$ of global functions on 
$\CM_{\rm H}(C_{0,n})$ to be the sub-algebra of $\mathfrak{A}_n$ generated by the 
functions commuting with the diagonal $\fsl_2$. The algebra $\CA$ contains the quantised 
Hamiltonians $\SH_r$, 
\begin{equation}\label{Hdef}
\mathsf{H}_r \equiv \sum_{s\neq r}\frac{\mathsf{J}_{rs}}{{z}_r-{z}_s}\,,
\qquad
\end{equation}
where the differential operator ${\CJ}_{rs}$ is defined as
\begin{equation}\label{kzmu}
\mathsf{J}_{rs}:=\eta_{aa'}{\SJ_r^a\SJ_s^{a'}} :=
{\SJ}^0_r {\SJ}^0_s + \frac{1}{2} ({\SJ}^+_r {\SJ}^-_s +{\SJ}^-_r {\SJ}^+_s)\,.
\end{equation}
The generators $\SH_r$ commute, $[\SH_r,\SH_s]=0$ for all $r,s$. Similar statements hold
for the conjugate algebra $\bar{\CA}$, which commutes with $\CA$ and contains 
the conjugate Hamiltonians $\bar{\SH}_r$, 
$r=1,\dots,n$.

A step towards the definition of suitable representations $\CR_n$ of $\mathfrak{A}_n$ 
is to choose a polarisation, a commutative sub-algebra of $\mathfrak{A}_n$ 
that will be represented by multiplication operators on $\CR_n$. In the present case there
are are two natural polarisations, defined by choosing either the sub-algbra
generated by $\sx_r$, $r=1,\dots,n$, or the one generated by
$\spp_r$, $r=1,\dots,n$. In both cases one gets an $n$-fold tensor product $\CR_n=\bigotimes_{r=1}^N \CP_n$
of representations $\CP_n$
of the Lie-algebra $\fsl_2$. In the first case one finds   a representation  realised by
 the differential operators
${\CJ}^\pm_r$, ${\CJ}^0_r$,
\begin{equation}\label{CJdef}
{\CJ}^-_r=\pa_{x_r} ,\quad
{\CJ}^0_r=x_r\pa_{x_r}-j_r,\quad
{\CJ}^+_r=-x^2_r\pa_{x_r}+2j_rx_r\,.
\end{equation}
The parameters $j_r$ appearing in \rf{CJdef} are related to the parameters 
$l_r$ of the classical Gaudin model by $l_r=-\ep_1j_r$.
In the polarisation generated by
$\spp_r$, $r=1,\dots,n$ we may choose the operators
\begin{equation}\label{tCJdef}
\tilde{\CJ}^-_r=p_r ,\quad
\tilde{\CJ}^0_r=-p_r\pa_{p_r},\quad
\tilde{\CJ}^+_r=-p_r\pa_{p_r}^2+\frac{j_r(j_r+1)}{p_r},
\end{equation}
as generators for the representation on $\CP_n$.
The Casimir operator is in both cases 
represented as multiplication by $j_r(j_r+1)$.

In order to fully define the relevant representations of the Lie algebra $\fsl_2$, one needs to 
specify the spaces of functions the differential operators defined in \rf{CJdef} and \rf{tCJdef}
should act on.
In the Gaudin model one usually considers finite-dimensional representations, 
restricting the choice of $j_r$ to $j_r=0,1/2,1, \dots$. The finite-dimensional representations 
can be realised via \rf{CJdef} on polynomial functions of the variables $x_r$.
We will mostly be interested in infinite-dimensional representations realised by means of the differential 
operators \rf{CJdef} on suitable spaces of non-polynomial functions. One may, for example, consider
representations defined by the differential operators $\CJ_r^a$ together with the 
conjugate operators $\bar{\CJ}_r^a$ obtained by $x_r\ra \bar{x}_r$, $\pa_{x_r}\ra \bar{\pa}_{\bar{x}_r}$
on certain (sub-)spaces of the space of smooth functions on $\BC$.  
The class of such representations contains the Lie algebra representations associated to 
principal series representations $\CP_n\equiv \CP_{j_n}$
of $\mathrm{SL}(2,\BC)$. The representations $\CP_{j_n}$ are unitary if $j_r\in -\frac{1}{2}+i\BR$.

The   symplectic quotient of $(T^*\mathbb{P}^1)^n$
by the three constraints \rf{reginfty} is naturally described by considering the action
of $\CA$ and $\bar{\CA}$ on the subspaces $\CR^{\rm inv}_n\subset \CR_n$ 
of invariants under the diagonal $\fsl_2$-action.
Representing the tensor product of representations $\CR_n=\bigotimes_{r=1}^N\CP_{j_n}$
in terms of functions $\Psi(\mathbf{x},\bar{\mathbf{x}})$ with $\mathbf{x}=(x_1,\dots, x_n)$
one may represent the elements of  $\CR_{\rm inv}$ as functions $\Psi(\mathbf{x},\bar{\mathbf{x}})$
which are invariant under the diagonal action of $\mathrm{SL}(2,\BC)$. We will find it more
convenient to represent the elements  of  $\CR_{\rm inv}$ as 
functions $\Psi(\mathbf{x},\bar{\mathbf{x}})$ of $n-1$ variables $\mathbf{x}=(x_1,\dots, x_{n-1})$
which are invariant under translations $x_r\ra x_r+b$ and 
behave under dilatations $x_r\ra a^2x_r$ as
\begin{equation}\label{scaling}
\Psi(a^2\mathbf{x},a^2\bar{\mathbf{x}})=a^{4J}\Psi(\mathbf{x},\bar{\mathbf{x}}), \qquad
J=-j_n+\sum_{r=1}^{n-1}j_r.
\end{equation}

The two representations \rf{CJdef} and \rf{tCJdef} are intertwined by the following slightly 
modified form of the Fourier-transformation.
\begin{align}
\Psi(\mathbf{x},\bar{\mathbf{x}})=\int d^2p_1\dots d^2p_{n-1} \;
\Phi(\mathbf{p},\bar{\mathbf{p}})
\,\prod_{r=1}^{n-1}
e^{p_r x_r-\bar{p}_r\bar{x}_r}
|p_r|^{-2j_r-2}.
\label{Fourier}
\end{align}
This map establishes an equivalence of the representation defined via \rf{CJdef} with 
a representation of the form \rf{tCJdef} 
in which a nilpotent generator is represented as multiplication operator.
We will refer to the representations defined on the functions $\Phi(\mathbf{p},\bar{\mathbf{p}})$
via \rf{tCJdef} as the Whittaker models for the 
representations $\bigotimes_{r=1}^{n-1}\CP_{j_r}$.  
One may note that the conjugate operators $\bar{\CJ}_r^\pm$, $\bar{\CJ}_r^0$ 
get mapped to the complex  conjugates of  $\tilde{\CJ}_r^\pm$, $\tilde{\CJ}_r^0$.

\subsection{Quantisation of Hitchin's Hamiltonians and the geometric Langlands correspondence} 
\label{geomLang}

Hitchin's Hamiltonians have been quantised in the work \cite{BD} of Beilinson and Drinfeld on the 
geometric Langlands correspondence. 
This means the following: There exist global differential operators $\SH_i$ on 
the line bundle $K^{1/2}$ on $\mathrm{Bun}_G$ such that the following holds:
\begin{itemize}
\item
The differential operators $\SH_i$ generate the commutative algebra $\mathfrak{D}$ of global 
differential operators acting on $K^{1/2}$, and
\item
the symbols of the differential operators $\SH_i$ coincide with generators of 
the algebra of functions on the Htichin base $\CB$ defined via Hitchin's map.
\end{itemize}
The construction in \cite{BD} uses elements of conformal 
field theory and the representation-theoretic results of \cite{FF92}. 
Our discussion follows the review \cite{Flect}.  

Beilinson and Drinfeld put the quantisation of the Hitchin in relation to 
the geometric Langlands correspondence, schematically represented as
\begin{equation}\label{rgeoLang}
\boxed{\;\; \;\phantom{\Big|}\text{${}^{\rm\sst L}_{}\mathfrak{g}$-opers}\;\;\;\;}
\quad\longleftrightarrow\quad
\boxed{\;\; \phantom{\Big|}
\CD-\text{modules on}\;\,\mathrm{Bun}_G
\;\;\;
}
\end{equation}
as we shall now briefly explain. The relation between the geometric Langlands correspondence and 
the Gaudin model was described in \cite{FFR}.

\subsubsection{Opers}


Opers are a special 
class of holomorphic connections $(\ep_1\pa_y+A(y))dy$  on $C$ 
with $A(y)$
being gauge equivalent to the form
$\big(\begin{smallmatrix} 0& t \\ 1 & 0\end{smallmatrix}\big)$. 
The equation defining horizontal sections $s$, $(\ep_1\pa_y+A(y)) s=0$, reduces 
to the ODE $(\ep_1^2\pa_u^2+t(u))s_2=0$ if 
$s=(\begin{smallmatrix}s_1\\s_2\end{smallmatrix})$. 
Covariance 
under changes of local coordinates
requires that $t=t(u)$ transforms  as 
\begin{equation}
{t}(u)=(y'(u))^2\,\tilde{t}_f(y(u))+\frac{\ep^2_1}{2}\{y,u\}\,,\quad \{y,u\}=
\left(\frac{y''}{y'}\right)'-\frac{1}{2}\left(\frac{y''}{y'}\right)^2, 
\end{equation}
identifying it as a {\it projective connection}. 
The underlying holomorphic bundle $\CE_{\rm op}$ must be an 
extension of the form $0\ra K^{\frac{1}{2}}\ra \CE_{\rm op}\ra K^{-\frac{1}{2}}\ra 0$. As $\CE_{\rm op}$
is uniquely defined thereby, an oper is completely specified by the choice of the projective connection $t$.

\subsubsection{Geometric Langlands correspondence}

One of the main results of 
Beilinson-Drinfeld is the existence of a canonical isomorphism 
of algebras 
\begin{equation}\label{BDthm}
\mathrm{Fun}\,\mathrm{Op}_{{}^{\sst \rm L}\fg}(C) \simeq \mathfrak{D}.
\end{equation}
This result implies a special case of the geometric Langlands correspondence.
Fixing an oper $\chi$ defines a homomorphism $\mathrm{Fun}\,\mathrm{Op}_{{}^{\sst \rm L}\fg}(C)\ra\BC$.
Using \rf{BDthm} one gets a homomorphism $\tilde{\chi}:\mathfrak{D}\ra\BC$. 
To each oper $\chi$ one
may assign a $\CD$-module $\Delta_\chi$ 
on $\mathrm{Bun}_G$ defined as
\begin{equation}
\mathfrak{D}_\chi=\mathfrak{D}/\mathrm{ker}\tilde{\chi}\cdot\mathfrak{D}.
\end{equation}
The correspondence between ${}^{\sst \rm L}\fg$-opers $\chi$ and $\CD$-modules $\mathfrak{D}_\chi$
on $\mathrm{Bun}_G$ constructed in this way is an  important 
part of what is called geometric Langlands correspondence.

This may be reformulated from the point of view of 
quantisation of the Hitchin system as follows:  
To an oper $\chi$
we may associate  the following system of differential equations on $\mathrm{Bun}_G$,
\begin{equation}\label{HitchEV}
\SH_if=E_i f,\qquad E_i=\tilde{\chi}(\SH_i).
\end{equation}
This system of differential equations is regular on the 
open dense subset of $\mathrm{Bun}_G$ containing the very stable
bundles, bundles that do not admit a nilpotent Higgs field. On this locus it defines a
vector bundle with  flat connection. Conjecturally, the vector bundle 
has regular singularities along the singular locus. Horizontal sections of the flat 
connection defined by the  equations \rf{HitchEV} will generically have
nontrivial monodromy around the singular loci.

Observing that the differential equations \rf{HitchEV} are the eigenvalue equations 
for Hitchin's Hamiltonians, it seems natural to interpret the results above
as the statement that $\mathrm{Op}_\mathfrak{{}^{\rm\sst L}\fg}(C)$ 
represents the natural geometric ``home'' for the eigenvalues of the quantised 
Hitchin Hamiltonians. The space of opers $\mathrm{Op}_{{}^{\rm L}\fg}(C)$ on 
$C$ represents the quantum analog $\CB_{\ep_1}$ of the base $\CB$ of the
Hitchin fibration.

\section{Quantum Separation of Variables}

We had noted in Section \ref{geomLang} that the geometric Langlands correspondence
is related to the eigenvalue problem of the quantised Hitchin Hamiltonians. It characterises 
the set of eigenvalues for which multi-valued analytic solutions can exist in terms of 
the opers associated to the  Lie algebra ${}^{\rm\sst L}\fg$. 
In all the cases where the Separation of Variables (SOV) approach 
has been developed it gives a concrete realisation of a correspondence 
between opers and eigenfunctions of the quantised Hitchin Hamiltonians. 
This has been fully realised when the surface $C$ has genus 
$g=0$ \cite{Ficmp} or $g=1$ \cite{EFR,FeS,HS} with any number
of punctures.
The SOV approach therefore offers an alternative approach to the geometric
Langlands correspondence which is similar to the first construction of such 
a correspondence due to Drinfeld \cite{Dr}, as has been pointed out in \cite{Ficmp}.
It  is natural to expect that the SOV approach can be extended to the cases with $g>1$,
furnishing a more concrete realisation of the geometric
Langlands correspondence in all cases.

In this section we will briefly describe how the SOV approach works in the case of genus zero, and then 
formulate a conjecture about the generalisation of the emerging picture to higher genus.

\subsection{Genus zero} 

The goal is to solve the eigenvalue problem
\begin{equation}
\SH_r\Psi_{\mathbf{E}}(\mathbf{x},\bar{\mathbf{x}})=E_r
\Psi_{\mathbf{E}}(\mathbf{x},\bar{\mathbf{x}}),\qquad
\bar{\SH}_r\Psi_{\mathbf{E}}(\mathbf{x},\bar{\mathbf{x}})
=\bar{E}_r\Psi_{\mathbf{E}}(\mathbf{x},\bar{\mathbf{x}}),
\end{equation}
where $\Psi_{\mathbf{E}}(\mathbf{x},\bar{\mathbf{x}})$ is a function of the 
$n-1$ variables $\mathbf{x}=(x_1,\dots, x_{n-1})$ and their complex conjugates
which are invariant under translations $x_r\ra x_r+b$ and behave 
under dilatations $x_r\ra a^2x_r$ as in \rf{scaling}.

The first step is to pass to the Whittaker model by means of the inverse of the Fourier-transformation
\rf{Fourier}, expressing solutions 
$\Psi_{\mathbf{E}}(\mathbf{x},\bar{\mathbf{x}})$
in terms of the 
eigenfunctions $\Phi_{\mathbf{E}}(\mathbf{p},\bar{\mathbf{p}})$
in the Whittaker model.
Let us then, following Sklyanin \cite{Skl},  perform the change of variables $\mathbf{p}\ra (u_0,\mathbf{u})$
defined by the family of equations
\begin{equation}
\vf_-(y)=\sum_{r=1}^{n-1}\frac{p_r}{y-z_r} =u_0\frac{\prod_{k=1}^{n-3}(y-u_k)}{\prod_{r=1}(y-z_r)}
\quad\Rightarrow\quad p_r(u)\,=\,u_0\frac{\prod_{k=1}^{n-3}(z_r-u_k)}{\prod_{s\neq r}^{n-1}
(z_r-z_s)}.
\end{equation}
Abusing notations we will denote  
$\Phi_{\mathbf{E}}(\mathbf{p}(u_0,\mathbf{u}),\bar{\mathbf{p}}(u_0,\mathbf{u}))$
by $\Phi_{\mathbf{E}}(\mathbf{u},\bar{\mathbf{u}})$.
Using identities like
\begin{equation}
\pa_{u_k}=\sum_{r=1}^{n-1}\frac{\pa p_r}{\pa u_k}\pa_{p_r}=
\sum_{r=1}^{n-1}\frac{1}{u_k-z_r}p_r\pa_{p_r},
\end{equation}
it becomes straightforward to 
show that the eigenvalue equation become equivalent to the 
set of ordinary differential equations 
\begin{equation}
(\ep_1^2\pa_{u_k}^2+t(u_k))\Phi_{\mathbf{E}}(\mathbf{u},\bar{\mathbf{u}})=0,\qquad
(\ep_1^2\bar{\pa}_{\bar{u}_k}^2+\bar{t}(\bar{u}_k))\Phi_{\mathbf{E}}(\mathbf{u},\bar{\mathbf{u}})=0,
\end{equation}
which can be solved in factorised from 
$\Phi_{\mathbf{E}}(\mathbf{u},\bar{\mathbf{u}})=\prod_{k=1}^{n-3}\phi_k(u_k,\bar{u}_k)$.
Further details can be found in \cite{Skl,Ficmp}.

The transformation from eigenfunctions $\Psi_{\mathbf{E}}(\mathbf{x},\bar{\mathbf{x}})$ to the 
functions $\Phi_{\mathbf{E}}(\mathbf{u},\bar{\mathbf{u}})$  
can be inverted explicitly \cite{FGT}. The inverse may be 
represented as an integral transformation of the form
\begin{align}
\Psi_{\mathbf{E}}(\mathbf{x},\bar{\mathbf{x}}) &\,=\,N_J\int d^2u_1\dots d^2u_{n-3}\;\mathcal{K}^{\rm\sst SOV}(x,u)\,
\Phi_{\mathbf{E}}(\mathbf{u},\bar{\mathbf{u}})\,,
\label{SOVtrsf}\end{align}
where the kernel $\mathcal{K}^{\rm\sst SOV}(x,u)$ can be represented explicitly as
\begin{align}\label{KSOVdef}
&\mathcal{K}^{\rm\sst SOV}(x,u)=
\left|\,{\sum_{r=1}^{n-1}x_r
\frac{\prod_{k=1}^{n-3}(z_r-u_k)}{\prod_{s\neq r}^{n-1}
(z_r-z_s)}}\right|^{2J}\,
\prod_{r=1}^{n-1}\Bigg|\frac{\prod_{s\neq r}^{n-1}
(z_r-z_s)}{\prod_{k=1}^{n-3}(z_r-u_k)}\Bigg|^{2(j_r+1)}\,
\prod_{k<l}^{n-3}|u_k-u_l|^{2}\,.
\end{align}
The integral transformation \rf{SOVtrsf} with kernel \rf{KSOVdef} is manifestly 
well-defined for  generic $(\mathbf{x},\bar{\mathbf{x}})$ when the 
real parts of the parameters $j_r$ are small enough. It may be defined for more 
general values of these parameters by analytic continuation.
Integrable singularities of a specific type 
occur at certain loci in the space parameterised by the variables $\mathbf{x}$.

\subsection{Higher genus}

The SOV approach appears to be less completely understood in the higher 
genus cases, but there is  evidence that the qualitative picture remains essentially 
unchanged \cite{Ficmp}. The first 
construction of the geometric Langlands correspondence due to Drinfeld \cite{Dr}
starts from a symmetric product $\CD$-module represented by an oper.
This $\CD$-module can be seen as 
the result of the canonical quantisation 
of the coordinates $(\mathbf{u},\mathbf{v})$ introduced in Section \ref{class}. Indeed, 
choosing a polarisation where the coordinates $u_k$ get represented as multiplication 
operators, and the coordinates $v_k$ as derivatives $\ep_1\pa_{u_k}$, one may 
identify the differential equations $(\ep_1^2\pa_{u_k}^2+t(u_k))\psi(u_k)=0$ as 
a quantum counterpart of the equation $v_k^2+\mathrm{tr}(\vf^2(u_k))=0$ 
defining the spectral curve $\Sigma$.

The description of 
Drinfeld's construction presented in \cite{Ficmp} reveals the striking similarity
of this construction with a quantum version of the SOV approach. It seems natural
to conjecture that the resulting $\CD$-modules on $\mathrm{Bun}_G(C)$ are
isomorphic to the ones furnished by the construction of Beilinson and Drinfeld.
The isomorphism of the $\CD$-modules provided by Drinfeld's first, and 
Beilinson and Drinfeld's second construction of the geometric Langlands correspondence
would imply the existence of a quantum version of the 
SOV for the Hitchin system \cite{Ficmp}.\footnote{The existence of such an
isomorphism would follow from the uniqueness of the irreducible Hecke
eigensheaf associated to an oper via the constructions in \cite{Dr} and \cite{BD}, 
which has not been established in the
literature, as far as we know. According to E. Frenkel, the
isomorphism can also be proved directly from the Hecke eigenvalue property. The author thanks
E. Frenkel for pointing this out to him.}

The resulting picture may be described a bit more concretely as follows.
Using extensions to represent the bundles $\CE$ allows us to introduce
$3g-2$ coordinates  $\mathbf{x}=(x_1,\dots, x_{3g-2})$ for the 
vector space $H^1(\CL^{-1})$ of extension classes. 
Eigenfunctions of the quantised Hitchin Hamiltonians may then be represented  
in terms of functions
$\Psi_{\mathbf{E}}(\mathbf{x},\bar{\mathbf{x}})$ of the coordinates $\mathbf{x}$ 
and their complex conjugates
which behave homogeneously  of weight $-3g+2$ under dilatations $x_r\ra a^2x_r$.
A Fourier-transformation similar to \rf{Fourier} will describe
the passage to the Whittaker model in which the quantum counterpart $\sf(y)$ 
of $\vf_-(y)$ is realised as a multiplication operator. The eigenvalue $f(y)$
of $\sf(y)$ is a section of the line bundle $\CL\otimes K_C$. It is then natural 
to introduce the zeros $u_k$ of $f(y)$ as new variables, and to rewrite the 
eigenvalue equations of the quantised Hitchin Hamiltonians in these variables.
The discussion above motivates the following conjecture: 
\begin{quote}
{\it The quantised Hitchin system admits a Whittaker model  representing 
eigenfunctions $\Phi_{\mathbf{E}}(\mathbf{u},\bar{\mathbf{u}})$ 
of the Hitchin Hamiltonians 
as sections of $(K^\frac{1}{2}\ot \bar{K}^\frac{1}{2})^{{3g-3}}$
which satisfy 
\begin{equation}\label{EV-SOV}
(\ep_1^2\pa_{u_k}^2+t(u_k))\Phi_{\mathbf{E}}(\mathbf{u},\bar{\mathbf{u}})=0,\qquad 
(\ep_1^2\bar{\pa}_{\bar{u}_k}^2+\bar{t}(\bar{u}_k))\Phi_{\mathbf{E}}(\mathbf{u},\bar{\mathbf{u}})=0,
\end{equation}
with $t(u)$ representing an oper on $C$.}
\end{quote}
This would represent a more concrete realisation of the geometric Langlands correspondence 
proven by Beilinson and Drinfeld. Note that no quantisation condition has been imposed yet. 
Inverting the steps outlined above should enable us to construct an integral transformation similar 
to \rf{SOVtrsf} allowing us to construct (generically multi-valued) eigenfunctions of Hitchin's
Hamiltonians from opers on $C$.





\section{Quantisation conditions}

The main questions to be addressed in this paper concern exactly what is bypassed in the 
geometric Langlands correspondence by the use of the $\CD$-module 
theory: What are interesting spaces of functions or sections of suitable line-bundles
in which one can search for the solutions of the eigenvalue equations for Hitchin's 
Hamiltonians, and given a particular choice, what are the possible eigenvalues?

This issue will be referred to as the choice of quantisation conditions. After identifying 
a choice of quantisation condition that appears to be particularly natural for the Hitchin 
system, we will explain how this choice can be reformulated as a condition on the monodromy
of the differential equations $(\ep_1^2\pa_u^2+t(u))\psi(u)=0$ representing the 
spectral problem in the SOV representation.

\subsection{Natural choices  of quantisation conditions}

It may be instructive
to compare the situation with the case of spectral problems for Schr\"odinger type operators
$H=-\pa_y^2+V(y)$,
for simplicity restricting attention to functions on the real line. For sufficiently regular
potentials there will exist two linearly independent solutions of the eigenvalue equations
$H\psi_E(y)=E\psi_E(y)$ for all real or even complex values of $E$. Physics usually
motivates us to impose additional requirement on the solutions $\psi_E(y)$ like
square-integrability which often can only be satisfied for a discrete set of values of $E$.
In other cases one may be interested in functions $\psi_E(y)$ which are periodic in $y$
which may again restrict the possible choices of $E$ to a discrete set. 
The supplementary conditions used to define the spectral problem of interest precisely 
will henceforth be referred to as quantisation conditions. Their mathematical content is to 
specify the exact class of functions which can be a solution 
to the spectral problem.

The Gaudin model is usually defined by considering functions of the variables $x_r$
appearing in the definition of the Gaudin-Hamiltonians which are polynomial in $x_r$
with degrees being given in terms of the parameters $j_r$ as $2j_r$. The representations 
defined on such functions via \rf{CJdef} correspond to the finite-dimensional representations of 
$\mathrm{SU}(2)$. This is one possible type of quantisation conditions defining what is 
called the $\mathrm{SU}(2)$ Gaudin model.

In this paper we are interested in  another type of quantisation condition. Following the discussion 
above we will consider the  pair of eigenvalue equations
\begin{equation}
\SH_r\Psi(\mathbf{x},\bar{\mathbf{x}})=E_r\Psi(\mathbf{x},\bar{\mathbf{x}}),\qquad
\bar{\SH}_r\Psi(\mathbf{x},\bar{\mathbf{x}})=\bar{E}_r\Psi(\mathbf{x},\bar{\mathbf{x}}),
\end{equation}
where $\overline{\SH}_r$ is the conjugate of $\SH_r$
obtained by replacing $x_r\ra \bar{x}_r$, $\pa_{x_r}\ra \bar{\pa}_{\bar{x}_r}$.
We are interested in solutions $\Psi(\mathbf{x},\bar{\mathbf{x}})$ which are
real-analytic away from possible singularities of the differential operators $\SH_r$ that
are furthermore {\it single-valued}. We had seen above that
such quantisation conditions are natural if one 
replaces the representations of $\mathrm{SU}(2)$ in the Gaudin model by principal series 
representations of $\mathrm{SL}(2,\BC)$. 

A similar type of quantisation condition can be considered for the quantum Hitchin system on 
higher genus surfaces $C$. The quantum Hitchin Hamiltonians are conjectured to have regular singularities
away from the locus within $\mathrm{Bun}_G(C)$ consisting of the very stable bundles.
For generic choice of an oper $\chi$, the solutions $f$ of \rf{HitchEV} will have nontrivial monodromies 
around the singular loci. For certain opers $\chi$ there may exist an hermitian 
form on the space of solutions which is invariant under the monodromy 
action, allowing us to construct single-valued solutions in 
the form of linear combinations of products of 
elements of a basis for the space of solutions to the 
eigenvalue equations multiplied by elements of the complex-conjugate basis.

Solutions to spectral problems establish generalised duality relations, 
in the simplest cases  relating certain 
spaces of functions $\mathfrak{S}$ to the spaces of functions on the sets of eigenvalues of 
commuting differential operators acting on functions in $\mathfrak{S}$.
Imposing additional conditions on one side will be reflected by
additional restriction occurring on the other side. 
From this point of view we may view the geometric Langlands correspondence as the solution 
to a natural pre-quantisation problem. It characterises the space dual 
to the multi-valued solutions of the Hitchin eigenvalue equations as the space of opers.
This sets the stage for the description of single-valued
solutions to the Hitchin eigenvalue equation to be proposed below.

From a physicist's perspective one might be tempted to look for a natural scalar product,
and to look for normalisable solutions within the class of single-valued ones. 
This would be a natural next step.
At the moment we have little to say about it. 

The reader may notice that the idea to combine holomorphic with anti-holomorphic 
functions into single-valued objects is familiar from conformal field theory. 
What we are proposing here is related to a particular limit of the so-called 
$H_3^+$-WZNW model \cite{RT,T10}.

When this paper was undergoing final revisions, E. Frenkel
pointed out to us that ideas similar to the ones presented above have been
discussed in the talk he gave at MSRI in Sept. 2014 \cite{Ftalk}.

\subsection{Quantisation versus classification of real projective structures}

The SOV transformation maps single-valued common eigenfunctions of the Hitchin 
Hamiltonians to single-value functions having the factorised form 
\begin{equation}\label{SOVrep}
\Phi_{\mathbf{E}}(\mathbf{u},\bar{\mathbf{u}})= \prod_{r=1}^h\phi_{\mathbf{E}}(u_r,\bar{u}_r)\,.
\end{equation}
Our goal is therefore to analyse the single-valuedness of the expression \rf{SOVrep}. 

As a preparation let us now  note that
each oper defines a projective structure on $C$, an atlas of local coordinates on $C$ with 
transition functions all represented as M\"obius transformations.  Indeed, given two linearly
independent solutions $\chi_i$ of  $(\ep_1^2\pa_u^2+t(u))\chi_i=0$, $i=1,2$ one may show
that $y(u)=\chi_1/\chi_2$ satisfies $2t(u)=\ep_1^2\{y,u\}$. Using  $y$ as a new local coordinate 
one therefore has $\tilde{t}(y)\equiv 0$. The M\"obius transformations are the
only allowed transition functions in an atlas formed by a collection of local charts with $\tilde{t}(y)\equiv 0$.
Background on projective structures relevant for us
is reviewed in \cite{Go04,Du}.

\subsubsection{Single-valuedness} 

In the form \rf{SOVrep} it becomes much easier to analyse the condition that
$\Phi_{\mathbf{E}}(\mathbf{u},\bar{\mathbf{u}})$ should be single-valued.
This will be the case iff the function $\phi_{\mathbf{E}}(u,\bar{u})$ which 
can be decomposed into linearly independent solutions of the differential 
equations \rf{EV-SOV} as
\begin{equation}\label{phians}
\phi_{\mathbf{E}}(u,\bar{u})=\sum_{i,j=1}^2C_{ij}\chi_i(u)\bar{\chi}_j(\bar{u}),
\end{equation}
has the property to be single valued. By a change of basis in the space
of solutions one may always bring the matrix $C_{ij}$ in \rf{phians} into diagonal form.
By a rescaling and multiplication of $C_{ij}$ by an overall phase one may assume that the 
diagonal matrix elements are contained in $\{-1,0,1\}$. The matrix $C_{ij}=\de_{ij}$ and diagonal 
matrices $C_{ij}$ having a vanishing diagonal matrix element are invariant under 
representations $\rho:\pi_1(C)\ra \mathrm{SL}(2,\BC)$ which never occur as holonomies of 
projective structures \cite{GKM}.
For the discussion of the remaining case $C=\mathrm{diag}(1,-1)$
we may may further transform $C_{ij}$ to the form
$C_{ij}=\ep_{ij}$, where $\ep_{12}=1$, $\ep_{ij}=-\ep_{ji}$, having invariance under
$\mathrm{SL}(2,\BR)$. Projective structures having holonomy in 
$\mathrm{PSL}(2,\BR)$ are called {\it real} projective structures.
It follows from the observations above that solutions to the single-valuedness condition correspond to
real projective structures.

We may thereby conclude that $\Phi_{\mathbf{E}}(\mathbf{u},\bar{\mathbf{u}})$ can be 
a single-valued solution to the system of equations \rf{EV-SOV}
only if the monodromy of $\ep_1^2\pa_u^2+t(u)$
 is conjugate to a homomorphism of $\pi_1(C)$
into  $\mathrm{SL}(2,\BR)$. The solution $\Phi_{\mathbf{E}}(\mathbf{u},\bar{\mathbf{u}})$ can then be 
represented in the factorised form \rf{SOVrep}.

One solution of the conditions above is well-known: For each Riemann surface $C$ there 
exists a unique metric $d^2s=e^{2\vf}dyd\bar{y}$ of constant negative curvature. 
The corresponding projective connection $t(y)=-\frac{1}{4}(\pa_y\vf)^2+\frac{1}{2}\pa_y^2\vf$ has the Fuchsian group 
$\Gamma$ uniformising $C\simeq \mathbb{H}/\Gamma$ as its holonomy. We may use it to construct a particular solution $\Phi_{\mathbf{E}_0}(u,\bar{u})$
to the quantisation 
conditions via \rf{SOVrep} and \rf{phians}. 
The function $\phi_{\mathbf{E}_0}^{}(u,\bar{u})$ appearing in the factorised representation \rf{SOVrep} 
for $\Phi_{\mathbf{E}_0}(u,\bar{u})$ is related to the metric of constant negative curvature 
as $\phi_{\mathbf{E}_0}^{}(u,\bar{u})=e^{-\vf(u,\bar{u})}$.

There exists a construction called grafting allowing to construct from a given 
projective structure with Fuchsian holonomy infinitely many other projective structures 
with the same holonomy \cite{Go87}. It was furthermore shown in \cite{Go87} that {\it all} projective structures
with holonomy being a fixed Fuchsian group are obtained by grafting the projective 
structure furnished by the uniformisation theorem, leading to a classification of the 
projective structures with Fuchsian holonomy.
Projective structures with Fuchsian holonomy not coming from the uniformisation of $C$ 
are called exotic. 
The  functions $\phi_{\mathbf{E}}(u,\bar{u})$ corresponding to exotic projective 
structures with Fuchsian holonomy define metrics of constant negative 
curvature via $e^{2\vf(u,\bar{u})}=(\phi_{\mathbf{E}_0}(u,\bar{u}))^{-2}$ only away from certain singular 
loci \cite{Ta14}.

There exist real projective structures not having Fuchsian holonomy. 
Such projective structures can be 
obtained by a small generalisation of the grafting construction
and have also been classified in \cite{Go87}.
A subclass of the real projective structures has holonomy in ${\rm PSL}(2,\BR)$.
In work in progress \cite{FT} we will describe in more detail which non-Fuchsian 
real projective structures can correspond to single-valued eigenfunctions of the 
quantised Hitchin Hamiltonians.

\begin{quote}
{\it The quantum Separation of Variables establishes a one-to-one correspondence between
single-valued eigenfunctions of the quantised Hamiltonians of the $\mathrm{SL}(2)$ 
Hitchin system and projective structures with holonomy in ${\rm PSL}(2,\BR)$ on $C$.}
\end{quote}


The classification of real 
projective structures from \cite{Go87} may therefore be used to classify the solutions 
of the quantisation conditions of the Hitchin system.


\subsection{Reformulation in terms of  complex Fenchel-Nielsen coordinates}

The Riemann-Hilbert correspondence between flat connections $\pa_y+A(y)$ and
representations $\rho:\pi_1(C)\ra G$ relates 
the moduli space $\CM_{\rm flat}(C)$ of flat
connections on $C$ to the so-called character
variety
$\CM_{\rm char}(C)={\rm Hom}(\pi_1(C),{\rm SL}(2,\BC))/{\rm SL}(2,\BC)$.
Useful sets of coordinates for
$\CM_{\rm flat}(C)$ are given by the trace functions
$L_{\ga}:=\operatorname{\rm tr}\rho(\ga)$ associated to 
simple closed curves $\ga$ on $C$. 

Minimal sets of trace functions that can be used to
parameterise $\CM_{\rm flat}(C)$ can be identified using
pants decompositions. Cutting a surface $C$ of genus $g$ with $n$ punctures
along a maximal set $\{\ga_1,\dots,\ga_{3g-3+n}\}$ 
on non-intersecting simple closed curves produces a surface having
connected components of type $C_{0,3}$ only.
Cutting $C$ along all but one of the 
curves in $\{\ga_1,\dots,\ga_{3g-3+n}\}$  produces a surface 
containing a single connected component of type $C_{0,4}$ or 
$C_{1,1}$. This component will be denoted as $C^{\dagger_r}$ if $\ga_r$
is the curve which was not cut. In order to get a coordinate system  for $\CM_{\rm char}(C)$
one needs two independent coordinates for each $C^{\dagger_r}$, $r=1,\dots,3g-3+n$. 
This is what we will define next.

\subsubsection{Complex Fenchel-Nielsen coordinates}

\begin{figure}[h]
\epsfxsize13.5cm
\centerline{\epsfbox{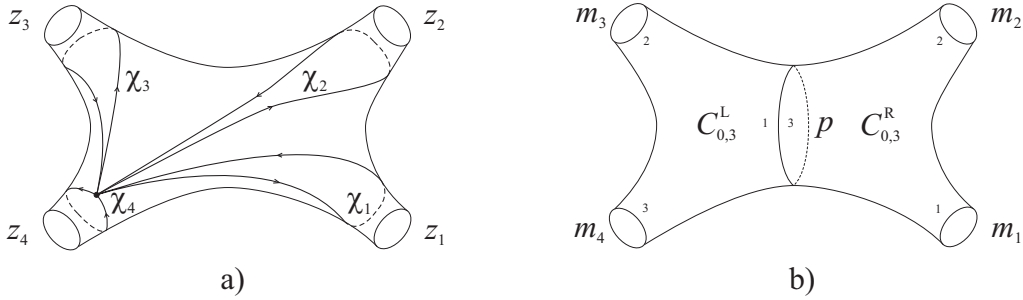}}
\caption{\it Basis of loops of $\pi_1(C_{0,4})$ and the decomposition $C_{0,4}=C_{0,3}^L\cup C_{0,3}^R$.}
\label{c04}\vspace{.3cm}
\end{figure}

Conjugacy classes of irreducible representations of $\pi_1(C_{0,4})$ are uniquely specified by
seven invariants
\begin{subequations}
\begin{align}\label{Mk}
&L_k=\operatorname{Tr} M_k=2\cos2\pi m_k,\qquad k=1,\ldots,4,\\
&L_s=\operatorname{Tr} M_1 M_2,\qquad L_t=\operatorname{Tr} M_1 M_3,\qquad L_u=\operatorname{Tr} M_2 M_3,
\end{align}
\end{subequations}
generating the algebra of invariant polynomial functions on $\CM_{\rm char}(C_{0,n})$. The monodromies $M_r$ are associated to the
curves $\chi_r$ depicted in Figure \ref{c04}. 
These trace functions satisfy the quartic equation
\begin{align}
 \label{JFR}
& L_1L_2L_3L_4+L_sL_tL_u+L_s^2+L_t^2+L_u^2+L_1^2+L_2^2+L_3^2+L_4^2=\\
 &\nonumber \quad=\left(L_1L_2+L_3L_4\right)L_s+\left(L_1L_3+L_2L_4\right)L_t
+\left(L_2L_3+L_1L_4\right)L_u+4.
\end{align}

The affine algebraic variety defined by (\ref{JFR}) is a concrete representation for
the character variety of $C_{0,4}$. For fixed choices of $m_1,\ldots,m_4$ in \rf{Mk}
one may use
equation \rf{JFR} to describe the character variety as a cubic surface in $\BC^3$.
This surface
admits a parameterisation in terms of coordinates $(\la,\kappa)$ of the form
\begin{subequations}\label{FN-FK}
\begin{align}
 L_s&=2\cos(\la/2),\\
   L_t\big((L_s)^2-4\big)&=
2(L_2L_3+L_1L_4)+L_s(L_1L_3+L_2L_4) \label{cl-'t Hooft}\\
& \quad
+2\cos(\kappa)
\sqrt{c_{12}(L_s)c_{34}(L_s)}\,,
\notag\\[1ex]
L_u\big((L_s)^2-4\big)&=
L_s(L_2L_3+L_1L_4)+2(L_1L_3+L_2L_4) \label{cl-dyonic}\\
& \quad
+2\cos((2\kappa-\lambda)/2)
\sqrt{c_{12}(L_s)c_{34}(L_s)}\,,
\notag
\end{align}
\end{subequations}
where $L_i=2\cos\frac{\la_i}{2}$, and $c_{ij}(L_s)$ is defined as
$c_{ij}(L_s)  =L_s^2+L_i^2+L_j^2+L_sL_iL_j-4$.

For $C\simeq C_{1,1}$ one may similarly parameterise the trace functions along the usual 
$a$- and $b$-cycles on the torus as
\begin{align}
& L_a=2\cos(\la/2)\,,\\
& L_b\big((L_a)^2-4\big)^{\frac{1}{2}}=2\cos(\kappa/2)
\sqrt{(L_a)^2+L_0-2},
\end{align}
with $L_0$ being the trace function associated to the boundary of $C_{1,1}$.

Using pants decompositions as described above one may define 
a pair of coordinates $(\kappa_r,\la_r)$ associated 
to each cutting curve $\ga_r$, $r=1,\dots,3g+n-3$.
Taken together, the tuples $\mathbf{k}=(\kappa_1,\dots,\kappa_{3g+n-3})$
and $\mathbf{l}=(\lambda_1,\dots,\la_{3g+n-3})$ form a system of coordinates for $\CM_{\rm flat}(C)$.
The coordinates defined above are Darboux coordinates,
\begin{equation}\label{ABform}
\Omega=\frac{1}{4\pi}
\sum_{r=1}^{3g-3+n}d\kappa_r\wedge d\lambda_r,
\end{equation}
where $\Omega$ is the symplectic form on the moduli spaces of flat 
connections introduced by Goldman and Atiyah-Bott, see \cite{Go04,NRS} and references therein.

\subsubsection{Quantisation conditions in terms of complex Fenchel-Nielsen coordinates}

We had previously reformulated the quantisation conditions as the condition that the
monodromy of the differential operator $\ep_1^2\pa_u^2+t(u)$ defines a representation of $\pi_1(C)$ in 
$\mathrm{PSL}(2,\BR)$. We are now going to reformulate this 
condition in terms of 
the coordinates
$(\mathbf{l},\mathbf{k})$. 

To this aim let us recall some basic facts reviewed in \cite{Du}.
Fixing an oper represented by a projective connection $t_*$ one may represent generic opers
via $t=t_*+\vartheta$, where $\vartheta$ is a quadratic differential.
Noting that $d:=\mathrm{dim} H^0(C,K^2)=3g-3+n$ we see that the opers represent a half-dimensional 
subspace $\mathrm{Op}_{\fsl_2}(C)$ of the moduli space of flat $\mathrm{SL}(2,\BC)$-connections. 
The monodromy map from $\mathrm{Op}_{\fsl_2}(C)$ to the character variety is locally 
biholomorphic. Fixing coordinates $\mathbf{E}=(E_1,\dots,E_{d})$ for the vector space $H^0(C,K^2)$
one may therefore use the monodromy map 
to define $\mathbf{l}=\mathbf{l}(\mathbf{E})$ and $\mathbf{k}=\mathbf{k}(\mathbf{E})$ 
by analytic continuation in $\mathbf{E}$.

Let us next observe that the coordinates 
$(\mathbf{l},\mathbf{k})$ as introduced above are closely related
to the classical Fenchel-Nielsen
coordinates for the Teichm\"uller component in the
character variety $\CM_{\rm char}^{\BR}(C)$
parameterising the Fuchsian groups appearing in the uniformization
of Riemann surfaces. The Fenchel-Nielsen coordinates $(l_r,k_r)$ are
given in terms of the coordinates $(\lambda_r,\kappa_r)$ simply 
as $l_r=\mathrm{i}\lambda_r$, and $k_r=\mathrm{i}\kappa_r$. 
The coordinates $(l_r,k_r)$ are real on Fuchsian groups.
Note that the coordinates $(\la_r,\kappa_r)$ and
$(\la_r+4\pi \,n_r,\kappa_r+2\pi \,\nu_r\,m_r)$ give the same values of the trace functions
iff $n_r,m_r\in\BZ$, 
and $\nu_r=1$ if $C^{\dagger_r}$ is of type $C_{0,4}$, while 
$\nu_r=2$ if $C^{\dagger_r}$ is of type $C_{1,1}$.

These observations lead us to reformulate the quantisation conditions for the Hitchin system as follows:
\begin{quote} 
{\it To each eigenfunction of the quantised Hamiltonians of the $\mathrm{SL}(2)$ Hitchin system 
on $C$ which is single-valued 
there exists a projective structure on $C$ 
having holonomy with complex FN coordinates $(\mathbf{k},\mathbf{l})=
(\mathbf{l}(\mathbf{E}),\mathbf{k}(\mathbf{E}))$ satisfying  
\begin{equation}\label{FN-qcond}
\mathrm{Re}(\la_r)=2\pi\,n_r,\quad \mathrm{Re}(\kappa_r)=\nu_r\,\pi\,m_r,\quad n_r,m_r\in\BZ,\;\, r=1,\dots,d,
\end{equation}
where $\nu_r$ was defined above, and $d:=3g-3+n$.
}
\end{quote}

We observe an interesting point: The definition of the coordinates $(\mathbf{l},\mathbf{k})$
depends on the choice of a pants decomposition. Changing the pants decomposition must 
relate the integer points defined in \rf{FN-qcond} to each other.

It should be interesting to relate the conditions in  \rf{FN-qcond} to Goldman's classification 
of projective structures with real holonomy in \cite{Go87}. The data classifying such 
projective structures are collections $\mu(\mathbf{M})$ of disjoint simple closed curves $\mu_i$ on $C$
with positive integer weights $M_i$ attached to them. We may represent these projective
structures in terms of the sum of a reference oper with a quadratic differential 
$\vartheta=\vartheta(\mathbf{E}_{\mu(\mathbf{M})})$, defining a discrete set of points in 
$\CB_{\ep_1}\simeq \mathrm{Op}_{\fsl_2}(C)$. In this regard it is suggestive
to note that the grafting operation relating different projective structures with Fuchsian 
holonomy 
can be represented in terms of the complex Hamiltonian twist flows generated 
by the complex length functions \cite{Go04}. 
What is not clear to us at the moment
is how exactly this operation gets represented in terms of the coordinates
$\mathbf{l}(\mathbf{E})$.

\section{Generating functions of varieties of opers}

It follows from the observations above that the quantisations conditions 
are naturally described using the space $\CP(S)$ of projective structure on a 
surface $S$.
A projective structure defines 
in particular a complex structure. The space of projective structures $\CP(S)$ 
on a surface $S$ is therefore
fibered over the moduli space $\CM(S)$  of complex structures on $S$.

The definition of the generating function $\CW$ given below will use the results
of \cite{Ka,Lo,BKN} comparing two natural holomorphic symplectic structures 
on $\CP(S)$. The first is defined using the symplectic structure \rf{ABform}
on the character variety via the holonomy map.
The second comes from the non-canonical isomorphisms of $\CP(S)$ to the cotangent  bundle 
$T^*\CM(S)$ briefly reviewed below, referring to \cite{Ka,Lo,BKN} for a more detailed
review of the relevant background and further references.

\subsection{Generating function $\CW$}

Let us first recall that the space of quadratic differentials on a Riemann surface $C$
is canonically isomorphic to the cotangent fiber 
$T^*\CM(S)$ at the point in the moduli space $\CM(S)$ represented by the Riemann surface $C$.  
This means that  $\CP(S)$ is non-canonically isomorphic to $T^*\CM(S)$. 
A set of local coordinates $\mathbf{q}=(q_1,\dots,q_d)$ for $\CM(S)$ canonically defines a set 
$(\mathbf{q},\mathbf{E})$ of 
coordinates for $T^*\CM(S)$, with $\mathbf{E}=(E_1,\dots,E_d)$ being the coefficients
in the expansion of $\vartheta\in H^0(C,K^2)$ with respect to the dual of the basis for the tangent space
$T\CM(S)$ generated by the vector fields 
$\pa_{q_r}$, $r=1,\dots,d$. The isomorphisms $\CP(S)\simeq T^*\CM(S)$ defined
by the choice of a reference projective connection allow us 
to use  $(\mathbf{q},\mathbf{E})$ as  coordinates for $\CP(S)$. 

Associating to the projective connection $t(y)$
the holonomy of the connection $(\pa_y+(\begin{smallmatrix} 0& t \\ 1 & 0\end{smallmatrix}))dy$ defines
a map from $\CP(S)$ to the character variety. 
It follows from the theorems proven in \cite{Ka,Lo,BKN} that this map relates the natural symplectic structures, 
\begin{equation}
\Omega=\frac{1}{\mathrm{i}}\sum_{r=1}^{h}dE_r\wedge dq_r=\frac{1}{4\pi}
\sum_{r=1}^{h}d\kappa_r\wedge d\lambda_r\,.
\end{equation}
The change of Darboux-coordinates 
from $(\mathbf{q},\mathbf{E})$ to $(\mathbf{l},\mathbf{k})$ 
can be described by a generating function $\CW(\mathbf{l}, \mathbf{q})$ satisfying
\begin{equation}\label{Wtok}
\kappa_r(\mathbf{l},\mathbf{q})
=-4\pi\mathrm{i}\frac{\pa}{\pa \la_r}\CW(\mathbf{l},\mathbf{q}),\qquad
\frac{\pa}{\pa q_r}\CW(\mathbf{l}, \mathbf{q})=E_r.
\end{equation}
It follows in particular that the subspaces of opers in the moduli spaces of flat connections are 
Lagrangian, and that the functions $\CW$ are the generating function of this Lagrangian subspace.
Note that the functions  $\CW(\mathbf{l},\mathbf{q})$ satisfying \rf{Wtok} depend on the choices of coordinates
$(\mathbf{q},\mathbf{E})$ and $(\mathbf{l},\mathbf{k})$. Equation \rf{Wtok} furthermore determines 
$\CW(\mathbf{l},\mathbf{q})$ only up to a constant. We will discuss the resulting issues
in Section \ref{Globaldef} below.

\subsection{Quantisation conditions in terms of the generating function $\CW$}


It remains to observe that the quantisation conditions can be reformulated in
terms of the function $\CW(\mathbf{l},\mathbf{q})$ in a way which resembles the use of 
the function introduced by Yang and Yang \cite{YY} for the description of the quantisation 
conditions in quantum integrable models soluble by the Bethe ansatz method. We immediately find from 
\rf{FN-qcond} and \rf{Wtok} that

\begin{quote}
{\it For each  single-valued eigenfunction of the quantised Hamiltonians of the $\mathrm{SL}(2)$ Hitchin system 
on $C$ there exist 
tuples $\mathbf{n}=(n_1,\dots,n_{d})$,
$\mathbf{m}=(m_1,\dots,m_{d})$ of integers and  
a solution $\mathbf{l}=\mathbf{l}(\mathbf{n},\mathbf{m})$ 
to the equations
\begin{equation}\label{W-qcond}
\mathrm{Re}(\la_r(\mathbf{n},\mathbf{m}))
=2\pi\,n_r,\quad \mathrm{Re}\bigg(\frac{\pa}{\pa \la_r}
\CY(\mathbf{l},\mathbf{q})\bigg)\bigg|_{\mathbf{l}=\mathbf{l}(\mathbf{n},\mathbf{m})}
=\nu_r\,\pi\,m_r,
\end{equation}
where $r=1,\dots,d$, and the function $\CY(\mathbf{l},\mathbf{q})$ is given as
$\CY(\mathbf{l},\mathbf{q})=4\pi \mathrm{i}\,\CW(\mathbf{l},\mathbf{q})$. 

For a given tuple 
$(\mathbf{n},\mathbf{m})$ of integers characterising a single-valued eigenfunction one gets the
corresponding eigenvalues $\mathbf{E}=(E_1,\dots,E_{d})$ as
\begin{equation}
E_r(\mathbf{n},\mathbf{m})
=\frac{\pa}{\pa q_r}\CW(\mathbf{l}(\mathbf{n},\mathbf{m}), \mathbf{q}).
\end{equation}
}\end{quote}
This had previously been observed in \cite{T10} for the $\mathrm{SL}(2,\BC)$-Gaudin model.
In a parallel development it has been proposed in \cite{NRS} that quantisation 
conditions for the Hitchin system are naturally formulated in terms
of the function $\CW$. The proposal above completes the proposal from \cite{NRS} by
establishing the precise relation between a particular quantisation condition for the 
quantised Hitchin system to a specific condition formulated in terms 
of the function $\CW$.


\section{Global definition of the Yang's function} \label{Globaldef}

The goal of this section is to clarify which global geometric object is locally 
represented by the function $\CW$.  One may note that
the local definition for the functions $\CW$ given above is 
sufficient for the goal to 
formulate the quantisation conditions for the Hitchin integrable system.
Readers only interested in this aspect can safely skip this section.
However, from a mathematical point of view it seems desirable 
to clarify if there is a globally defined geometric object on $\CP(S)$  locally
represented by the functions  $\CW(\mathbf{l},\mathbf{q})$. 
We are now going to propose that one can define  line-bundles on
$\CP(S)$ having  $\CW(\mathbf{l},\mathbf{q})$ as their local sections.
The proposal can be motivated using the observation that
$\CW(\mathbf{l},\mathbf{q})$ represents the leading asymptotics of the Virasoro conformal
blocks for large central charge. It then follows from known facts on the conformal blocks \cite{TV}.

\subsection{Global issues in the definition of the functions $\CW$}

In order to define the coordinates $(\mathbf{q},\mathbf{E})$ one needs coordinates $\mathbf{q}_\imath$
defined in open sets $\CU_\imath\subset\CM(S)$ such that 
$\{\CU_{\imath};\imath\in\CJ\}$ forms a cover of $\CM(S)$, and a family of reference opers $t_\imath=
t_\imath(\mathbf{q}_\imath)$
holomorphic on $\CU_\imath$. Coordinates $(\mathbf{q}_\imath,\mathbf{E}_\imath)$ and 
$(\mathbf{q}_\jmath,\mathbf{E}_\jmath)$
defined on sets $\CU_\imath$ and $\CU_\jmath$ with nontrivial intersection 
$\CU_{\imath\jmath}=\CU_\imath\cap\CU_\jmath$ will transform as
\begin{equation}
\sum_{r=1}^{d}E_r^\imath \,dq_r^\imath=\sum_{r=1}^{{d}}E_r^\jmath \,dq_r^\jmath+
df_{\imath\jmath}(\mathbf{q}_\imath),
\end{equation}
with $f_{\imath\jmath}$ being locally defined functions on $\CU_{\imath\jmath}$. 
The functions $f_{\imath\jmath}$, being defined in this way only up to a constant, must satisfy 
the condition 
\begin{equation}\label{cycle1}
f_{\imath_\1\imath_\2}+f_{\imath_\2\imath_\3}=f_{\imath_\1\imath_\3}+
\phi_{\imath_\1\imath_\2\imath_\3},
\end{equation}
with  $\phi_{\imath_\1\imath_\2\imath_\3}$ constant 
on triple overlaps $ \CU_{\imath_\1}\cap\CU_{\imath_2}\cap\CU_{\imath_3}$. 
A collection of functions $f_{\imath\jmath}$ defined on the overlaps of a cover defines
what was called a {\it projective} line bundle in \cite{FS}.

The dependence  of the coordinates $(\mathbf{l},\mathbf{k})$  on the choice of 
a pants decomposition $\si$ will be made explicit by using the notation $(\mathbf{l}_\si,\mathbf{k}_\si)$.
The generating function for the change of coordinates from $(\mathbf{l}_\si,\mathbf{k}_\si)$
to $(\mathbf{q}_\imath,\mathbf{E}_\imath)$ will be denoted as $\CW_{\si\imath}(\mathbf{l}_\si,\mathbf{q}_\imath)$.

Changes of the defining coordinate systems are described as follows.
A change from coordinates $(\mathbf{l}_{\si_\1},\mathbf{k}_{\si_\1})$ to 
$(\mathbf{l}_{\si_\2},\mathbf{k}_{\si_\2})$ is described by a generating function 
$F_{\si_\1\si_\2}(\mathbf{l}_{\si_\1},\mathbf{l}_{\si_\2})$ such that
\begin{equation}
\sum_{r=1}^d (\kappa^{\si_\1}_rd\lambda^{\si_\1}_r-\kappa^{\si_\2}_rd\lambda^{\si_\2}_r)=
dF_{\si_\1\si_\2}(\mathbf{l}_{\si_\1},\mathbf{l}_{\si_\2}).
\end{equation}
The generating functions must satisfy a condition of the form
\begin{equation}\label{cycle2}
F_{\si_\1\si_\2}(\mathbf{l}_{\si_\1},\mathbf{l}_{\si_\2}(\mathbf{l}_{\si_\1},\mathbf{l}_{\si_\3}))
+
F_{\si_\2\si_\3}(\mathbf{l}_{\si_\2}(\mathbf{l}_{\si_\1},\mathbf{l}_{\si_\3}),\mathbf{l}_{\si_\3})
=
F_{\si_\1\si_\3}(\mathbf{l}_{\si_\1},\mathbf{l}_{\si_\3})+\Phi_{\si_\1\si_\2\si_3},
\end{equation}
with $\Phi_{\si_\1\si_\2\si_3}$ being constant.

The generating functions transform under changes of coordinates for $\CP(S)$ as
\begin{subequations}
\begin{align}
&\CW_{\si\imath_\1}(\mathbf{l}_\si,\mathbf{q}_{\imath_\1})
=
\CW_{\si\imath_\2}(\mathbf{l}_\si,\mathbf{q}_{\imath_\2}(\mathbf{q}_{\imath_\1}))
+f_{\imath_\1\imath_\2}(\mathbf{q}_{\imath_\1}),\\
&\CW_{\si_\1\imath}(\mathbf{l}_{\si_\1},\mathbf{q}_\imath)
=\CW_{\si_\2\imath}(\mathbf{l}_{\si_\2}(\mathbf{l}_{\si_\1},\mathbf{q}_{\imath}),\mathbf{q}_\imath)
+F_{\si_\1\si_\2}(\mathbf{l}_{\si_\1},\mathbf{l}_{\si_\2}(\mathbf{l}_{\si_\1},\mathbf{q}_{\imath})).
\end{align}
\end{subequations}
It is known that the $\phi_{\imath_\1\imath_\2\imath_\3}$ define a non-trivial cohomology 
class \cite{FS}. It is therefore not yet clear if there can be any globally defined object
locally represented by the functions $\CW$ on $\CP(S)$. 

\subsection{Use of the gluing construction}

However, there is a way out. There is a well-known construction 
defining Riemann surfaces of arbitrary topology by gluing three-punctured spheres. 
The gluing  construction identifies parameterised annular neighbourhoods of two punctures 
on a possibly disconnected Riemann surface $\check{C}$ to produce a new surface $C'$.
For each cutting curve $\ga_r$, $r=1,\dots,3g-3+n$, defining a pants decomposition one may 
introduce a parameter $q_r$ specifying the identification in the gluing construction 
in such a way that $q_r\ra 0$ corresponds to the nodal degeneration where the length of 
$\ga_r$ vanishes. In this way one gets
 families of coordinates $\mathbf{q}_\si$ for a neighbourhood $\CU_\si$ 
of the boundary component
in the moduli space $\overline{\CM}(S)$ associated to a pants decomposition $\si$. 
By varying the 
pants decompositions $\si$ one gets a cover of $\overline{\CM}(S)$.

It is possible to choose the identification maps in such a way that all transition functions
in the Riemann surface produced by the gluing construction are M\"obius transformations. 
This means that the Riemann surfaces defined in this way come equipped 
with a natural projective structure. We may use this projective structure to define the 
coordinates $E_r$. In this way we get a family of 
coordinate systems $(\mathbf{q}_\si,\mathbf{E}_\si)$ covering $\CP(S)$.

We may then consider the functions $\CW_\si(\mathbf{l}_\si, \mathbf{q}_\si)$ defined as the 
generating function for the change of coordinates  from $(\mathbf{l}_\si,\mathbf{k}_\si)$ to 
$(\mathbf{q}_\si,\mathbf{E}_\si)$. The functions $\CW_\si(\mathbf{l}_\si, \mathbf{q}_\si)$
transform under the changes of the coordinates $(\mathbf{l}_\si,\mathbf{k}_\si)$ and
$(\mathbf{q}_\si,\mathbf{E}_\si)$ as
\begin{equation}\label{moves}
\CW_{\si_\1}(\mathbf{l}_{\si_\1},\mathbf{q}_{\si_\1})
=f_{\si_\1\si_\2}(\mathbf{q}_{\si_\1})\!+\!
F_{\si_\1\si_\2}(\mathbf{l}_{\si_\1},\mathbf{l}_{\si_\2}\!(\mathbf{l}_{\si_\1},\!\mathbf{q}_{\si_\1}))\!+\!
\CW_{\si_\2}(\mathbf{l}_{\si_\2}\!(\mathbf{l}_{\si_\1},\!\mathbf{q}_{\si_\1}),\mathbf{q}_{\si_\2}(\mathbf{q}_{\si_\1})).
\end{equation}

We are now going to argue
that there exists a line-bundle over $\CP(S)$ having the functions 
 $\CW_\si(\mathbf{l}_\si, \mathbf{q}_\si)$ as its local sections. 
Indeed, our claim must hold if we are able to give an unambiguous
definition of the function $\CW_\si(\mathbf{l}_\si, \mathbf{q}_\si)$ 
satisfying \rf{Wtok} for the coordinates 
$(\mathbf{q}_\si,\mathbf{E}_\si)$ and $(\mathbf{l}_\si,\mathbf{k}_\si)$
associated to any pants decomposition $\si$.
In overlaps of the respective domains of definition 
there exist relations between $(\mathbf{q}_{\si_\1},\mathbf{E}_{\si_\1})$ and
$(\mathbf{q}_{\si_\2},\mathbf{E}_{\si_\2})$, as well as
$(\mathbf{l}_{\si_\1},\mathbf{k}_{\si_\1})$ and $(\mathbf{l}_{\si_\2},\mathbf{k}_{\si_\2})$. It follows that
there must exist  relations of the form \rf{moves}.
The consistency of these relations on triple overlaps then implies a cancellation between the constants
$\phi_{\imath_\1\imath_\2\imath_\3}$ and $\Phi_{\imath_\1\imath_\2\imath_\3}$ appearing
in the consistency conditions \rf{cycle1} and \rf{cycle2} for the generating functions $ f_{\si_\1\si_\2}$
and $F_{\si_\1\si_\2}$, respectively. 

One way to define the functions $\CW_\si(\mathbf{l}_\si, \mathbf{q}_\si)$ 
unambiguously is to  specify the asymptotic behaviour they have
at the maximal nodal degeneration of $C$ in the Deligne-Mumford compactification of $\CM(S)$. 
We claim that the following choice does the job:
\newcommand{\upc}{\Upsilon_{\rm\sst cl}}
\begin{align}
\CW_\si(\mathbf{l}_\si, \mathbf{q}_\si)\sim & \sum_{r\in I_{0,4}} ({\de(l_r)-\de(l_{r,1})-\de(l_{r,2})})\log q_r
+\sum_{r\in I_{1,1}} {\de(l_r)}\log q_r\\
& 
+\sum_{v\in \mathbb{P}_\si}N(l_{v,1},l_{v,2},l_{v,3})+\CO(q_r),
\notag\end{align}
where $I_{0,4}$ and $I_{1,1}$ are the subsets of $\{1,\dots,3g-3+n\}$ for which $C^{\dagger_r}\simeq C_{0,4}$ 
and $C^{\dagger_r}\simeq C_{1,1}$, respectively, 
$\mathbb{P}_\si$ is the set of pairs of pants appearing in the pants decomposition $\si$
of $C$, $l_{v,i}$, $i=1,2,3$, are the complex length coordinates associated to the boundary 
curves of the pair of pants labelled by $v\in\mathbb{P}_\si$, and 
$N(l_\3,l_\2,l_\1)$ is defined as 
\begin{align}
N(l_\3,l_\2,l_\1)=&\frac{1}{2}
\sum_{s_1,s_2=\pm}\upc\big(\fr{1}{2}+\fr{i}{4\pi}(s_\1l_\1+s_\2l_\2+l_\3)\big)-
\frac{1}{2}\sum_{i=1}^3\mathrm{Re}\big(\upc\big(1+\fr{i}{2\pi}l_i\big)\big)
\end{align}
assuming $l_i\in\BR$ to simplify the expression and using the notation
$
 \upc(x)=\int_{1/2}^{x}du\;\log\frac{\Gamma(u)}{\Gamma(1-u)}
$.
The proof of this claim is outlined in Section 10.3 and Appendix E of \cite{TV}. It is interesting to note
that the function $N(l_\3,l_\2,l_\1)$ appears in the semiclassical limit of the Liouville three point 
functions \cite{ZZ}.

\section{Concluding remarks}

\subsection{Semiclassical limit}\label{WKB}

The semiclassical limit of the quantisation conditions formulated above
is closely related to the geometric picture presented in Section \ref{class}.

In order to see this, we need to observe that the leading behaviour of the 
solutions to the differential equations $(\ep_1^2\pa_u^2+t(u))\chi(u)=0$ 
for $\ep_1\ra 0$ may be represented as
\begin{equation}
\chi(u)\sim e^{\frac{\mathrm{i}}{\ep_1}\int^udu' \,v(u')},
\end{equation}
where $v(u)$ satisfies $(v(u))^2=t(u)$. The asymptotic behaviour of the 
monodromies of the differential equations $(\ep_1^2\pa_u^2+t(u))\chi(u)=0$ 
can be expressed in terms of the periods of the canonical one-form 
$\la =vdu$ on the double cover $\Sigma=\{(u,v);v^2=t(u)\}\subset T^*C$,
\begin{equation}
L_{r,s}\sim \exp\big({\fr{\mathrm{i}}{\ep_1}a^{r}}\big),\qquad
L_{r,u}\sim \exp\big({\fr{\mathrm{i}}{\ep_1}a_{r}^{\rm\sst D}}\big),
\end{equation}
where $a^r$ and $a_r^{\rm\sst D}$ are the periods defined in 
\rf{periods} by integrating along a suitable  canonical basis for the first odd homology of  $\Sigma$.
It follows immediately from \rf{Wtok} that 
\begin{equation}
\CW\big(\fr{1}{\ep_1}\mathbf{a},\mathbf{q}\big)\sim\frac{1}{\ep_1}\CF(\mathbf{a},\mathbf{q})+
\mathrm{regular},
\end{equation}
where $\CF(\mathbf{a},\mathbf{q})$ is the potential appearing in the 
discussion of the algebraic integrability of the Hitchin system in Section \ref{special}. 
The coordinates $(\mathbf{a},\mathbf{a}^{\rm\sst D})$ form yet another set of Darboux coordinates 
for $T^*\CM(S)$
called homological coordinates in \cite{BKN}, and $\CF(\mathbf{a},\mathbf{q})$ is the 
generating function for the change of coordinates from $(\mathbf{a},\mathbf{a}^{\rm\sst D})$
to $(\mathbf{q},\mathbf{E})$.

The quantisation conditions \rf{W-qcond} therefore have the following leading asymptotics
\begin{equation}\label{BS}
\mathrm{Re}(a^r)=\ep_1\pi \, n_r,\qquad
\mathrm{Re}(a^{\rm\sst D}_r)=\ep_1 \pi \,m_r.
\end{equation}
These are the natural Bohr-Sommerfeld quantisation conditions for the real action-angle variables
introduced in Section \ref{special}, indicating that the quantisation conditions studied in this
paper are indeed very natural.

\subsection{Real versus complex integrable systems}

Somewhat different types of conditions expressed 
in terms of generalised Yang's functions have previously been 
found in other cases admitting such a  formulation \cite{NS}. Rather than the pair of 
real equations \rf {W-qcond} it was shown in \cite{NS} that the quantisation 
conditions for the Toda chain and for the elliptic Calogero-Moser models
can be represented as a single complex equation of the form
\begin{equation}\label{A-type}
\pa_{a^r}\CY(\mathbf{a},\mathbf{q})=2\pi n_r, \qquad r=1,\dots,d.
\end{equation}

It seems that the quantisation conditions of the type \rf{A-type} are natural in algebraically integrable
systems which are complexifications of {\it real} integrable systems like the Toda chain. 
In this paper we
have been considering integrable systems which are genuinely complex. The two 
types of quantisation conditions, \rf{A-type} and \rf{W-qcond} are naturally associated 
to these two cases, respectively, as is also supported by the semiclassical 
considerations in Section \ref{WKB} above. Comparing the results of a semiclassical analysis 
of the quantisation conditions for XXX-type spin-chains with $\mathrm{SL}(2,\BC)$-symmetry carried out in \cite{DKM}
with \rf{BS} indicates that quantisation conditions of such spin chains are of the same type as found for 
the Hitchin system in this paper.

\subsection{Relation to conformal field theory}

WZW-type conformal field theories can be defined mathematically using the representation 
theory of the affine Lie algebra $\hfg_k$ at level $k$ extending a semisimple finite-dimensional 
Lie algebra $\fg$.
The conformal blocks of WZW-type conformal field theories are defined as
elements $f$ in the dual of the vacuum representation $V_0$ of $\hfg_k$ invariant under the 
natural action of the Lie-algebra of meromorphic functions allowed to have poles 
only at a single point $P\in C$. The defining invariance condition may be 
twisted by families of  holomorphic $G$-bundles $\CE_\mathbf{x}$, introducing a 
dependence on a collection of parameters $\mathbf{x}=(x_1,\dots,x_d)$ representing 
coordinates on $\mathrm{Bun}_G$.

The conformal blocks 
can be characterised 
in terms of the solutions $\CZ(\mathbf{x},\mathbf{q})$ to the KZB-equations,
taking the form 
\begin{equation}\label{KZ}
(k+h^{\vee})\frac{\pa}{\pa q_r}\CZ(\mathbf{x},\mathbf{q})=\SH_r\,\CZ(\mathbf{x},\mathbf{q}),
\end{equation}
where $\mathbf{q}$ are complex coordinates for the Teichm\"uller space $\CT(S)$, 
$\SH_r$ are the quantised Hitchin-Hamiltonians and $h^\vee$ is the dual Coxeter number of $\fg$.
In the critical level limit $\ep_2\ra 0$, $\ep_2:=-(k+h^{\vee})\ep_1$ 
one may solve \rf{KZ} with the ansatz \cite{RV} (see \cite{BF} for related results)
\begin{equation}\label{WZWasym} 
\CZ(\mathbf{x},\mathbf{q})\sim e^{-\frac{\ep_1}{\ep_2}\CS(\mathbf{q};\ep_1)}
\Psi(\mathbf{x},\mathbf{q};\ep_1)(1+\CO(q)),
\end{equation}
where $\CS(\mathbf{q};\ep_1)$ and $\Psi(\mathbf{x},\mathbf{q};\ep_1)$ satisfy
\begin{align}
&\SH_r\Psi(\mathbf{x},\mathbf{q};\ep_1)=E_r\Psi(\mathbf{x},\mathbf{q};\ep_1),\qquad
\frac{\pa}{\pa {q_r}}\CS(\mathbf{q};\ep_1)=E_r.
\end{align}

This result can be made more precise  by using the gluing construction to construct bases 
of conformal blocks associated to pants decompositions. In the case $\fg=\fsl_2$ one gets solutions 
$\CZ(\mathbf{l};\mathbf{x},\mathbf{q})$
to the the KZB-equations depending on additional parameters $\mathbf{l}=(\la_1,\dots,\la_{d})$
parameterising
 the intermediate representations\footnote{The parameters $\la_r$ parameterise the weights 
$j_r$ of the intermediate representations as $j_r=-\frac{1}{2}+\mathrm{i}\la_r$.}  used in the gluing construction, 
one complex number $\la_r$ for each cutting curve $\ga_r$. The analysis 
above can then be refined by using the Verlinde loop operators in a similar way 
as it was done in \cite{FGT,T17b}, giving 
\begin{equation}\label{Stok}
\kappa_r(\mathbf{l},\mathbf{q})
=-4\pi\mathrm{i}\frac{\pa}{\pa \la_r}\CS(\mathbf{l},\mathbf{q}).
\end{equation}
It follows that  the function $\CS(\mathbf{l},\mathbf{q})$ representing the leading term in the asymptotics 
\rf{WZWasym} coincides with the function $\CW(\mathbf{l},\mathbf{q})$ 
studied in this paper.

\subsection{Real geometric Langlands}

The results of this note can be re-interpreted as a variant of the geometric Langlands program.
One aspect of the 
ordinary Langlands program is the classification of the (cuspidal) spectrum of the  Laplacian on certain 
locally symmetric spaces. From the point of view of integrable systems one may view the Laplacian
as the ``local'' observable one is interested in. Possible degeneracies can be resolved by using additional 
``non-local'' observables called Hecke operators. 

From this point of view one may interpret the geometric Langlands program as a conjectural answer to 
a ``pre-spectral'' problem. It describes the natural geometric home for the eigenvalues of the 
Hitchin Hamiltonians -  the variety of opers within the moduli space of local systems.

The natural next step is to define natural quantisation conditions defining what might be called 
cuspidal eigenfunctions of the Hitchin Hamiltonians.  In this paper we propose such a quantisation 
condition. It selects a discrete subset within the variety of opers - a particular subset of 
the intersections between the 
variety of opers and the real slice. We propose to view correspondences between
real opers and single-valued eigenfunctions of the Hitchin Hamiltonians as natural variants 
of the geometric Langlands correspondence.
\begin{equation}\label{rgeoLang2}
\boxed{\;\; \phantom{\Big|}
\text{Real ${}^{\rm L}_{}\mathfrak{g}$-opers}
\quad}
\quad\longleftrightarrow\quad
\boxed{\;\; 
\begin{aligned}&\text{Single-valued eigenfunctions}\\
&\text{of the Hitchin-Hamiltonians}
\end{aligned}\;\;
}
\end{equation}
As opposed to the versions of the geometric Langlands correspondence intensively 
studied in the literature, this version is not of algebro-geometric nature: It is based on the 
relation between the two natural algebraic structures on the moduli of flat connections 
furnished by the 
non-algebraic Riemann-Hilbert correspondence.
However, the version of the geometric Langlands correspondence proposed above 
has the virtue to be somewhat 
closer analogous to the original Langlands program, in the sense that the 
single-valued eigenfunctions of the Hitchin-Hamiltonians can be viewed as 
analogs of the automorphic forms. 

{\bf Acknowledgements.} 
The author would like to thank Leon Takhtajan and especially E. Frenkel for 
very useful comments on a previous version of the draft.

This work was supported by the Deutsche Forschungsgemeinschaft (DFG) through the 
collaborative Research Centre SFB 676 ``Particles, Strings and the Early Universe'', project 
A10.


\bibliographystyle{JHEP_TD}
\bibliography{Yang-Hitchin}

\end{document}